\documentstyle[preprint,aps,pre]{revtex}
\begin{document}
\tighten
\title{Rheological properties in a low-density granular mixture}
\author{Jos\'e Mar\'{\i}a Montanero\footnote[1]{Electronic address:
jmm@unex.es}}
\address{Departamento de Electr\'onica e Ingenier\'{\i}a Electromec\'anica,\\
Universidad de Extremadura, E-06071 Badajoz, Spain}
\author{Vicente Garz\'{o}\footnote[2]{Electronic address: vicenteg@unex.es}}
\address{Departamento de F\'{\i}sica, Universidad de Extremadura, E-06071 \\
Badajoz, Spain}

\date{\today}
\maketitle

\begin{abstract}

Steady simple shear flow of a low-density binary mixture of inelastic smooth
hard spheres is studied in the context of the Boltzmann equation. This equation 
is solved by using two different and complementary approaches: a Sonine 
polynomial expansion and the direct simulation Monte Carlo method. The 
dependence of the shear and normal stresses as well as of the steady 
granular temperature on both the dissipation and the parameters of the 
mixture (ratios of masses, concentration, and sizes) is analyzed. In 
contrast to previous studies, the theory predicts and the simulation 
confirms that the partial temperatures of each species are different, even 
in the weak dissipation limit. In addition, the simulation shows that the 
theory reproduces fairly well the values of the shear stress and the 
phenomenon of normal stress differences. On the other hand, here
we are mainly interested in analyzing transport in the homogeneous
shear flow so that, the possible formation of particle clusters is ignored in 
our description.

 
\end{abstract}
\draft
\vspace{1cm}

\bigskip \narrowtext

\section{Introduction}
\label{sec1}

Many features associated with dissipation in rapid granular flows can be 
well represented by a fluid of hard spheres with inelastic collisions. 
In the simplest model the grains are taken to be smooth so that the 
inelasticity is characterized by means of a constant coefficient of normal 
restitution. The essential difference with respect to normal fluids is 
the absence of energy conservation, which leads to modifications 
of the usual hydrodynamic equations.  In recent years, the Boltzmann and 
Enskog equations have been generalized to account for inelastic binary  
collisions. These equations have been solved by means of an expansion akin 
to the Chapman-Enskog method up to the Navier-Stokes order and detailed 
expressions for the corresponding transport coefficients have been obtained 
\cite{BDKS98,RET99}. These expressions are not restricted to the 
low-dissipation limit and comparison with Monte Carlo simulations indicate 
that the results are very accurate, even for strong 
dissipation \cite{BMC99}. In the context of multicomponent granular gases, 
most of the existing work appears to be based on weak dissipation 
approximations \cite{JM89,Z95,AW98,WA99}. Given that the inelasticity is
small, an usual assumption in these studies is to consider
a single temperature variable characterizing the entire mixture.
However, as one of the authors pointed out\cite{Z95}, the equipartition of
energy is not completely justified beyond the low-dissipation limit and it
is necessary to offer theories involving mixtures of granular materials in
which the kinetic temperatures of species $T_i$ are different from the
mixture temperature $T$. As a matter of fact, recent experiments
\cite{FM02} and simulations \cite{CH02} on {\em driven} granular mixtures
show that the two types of grains do not attain the same granular
temperature. In terms of the mean square velocities of species, this
implies a violation of the classical equipartition theorem.
A related finding for a binary mixture undergoing homogeneous
cooling (i.e., an unforced system) has been reported by
Garz\'o and Dufty \cite{GD99,GD01} from a kinetic theory analysis.


All the above works refer to near equilibrium situations. Very little is 
known about far from equilibrium states. This is true for both molecular 
and granular fluids due to the intricacy of the Boltzmann and 
Enskog collision operators. Nevertheless, the difficulties are even harder 
for granular gases since gradients in the system can be controlled by 
dissipation in collisions and not only by the boundary and initial 
conditions. Thus, for instance, a granular system with uniform boundaries 
at constant temperature develops spatial inhomogeneities \cite{BC98}.

One of the simplest far from equilibrium physical situations 
corresponds to the simple shear flow. Macroscopically, it is characterized 
by uniform density and temperature and a constant velocity profile. In the 
case of molecular fluids, this state is not stationary since the 
temperature increases monotonically in time due to viscous heating. However, 
for granular fluids a steady state is possible when the viscous heating is 
exactly balanced by the inelastic cooling. As a consequence, for a 
given shear rate, the temperature is a function of the restitution 
coefficient in the steady state. This steady state is precisely what we want 
to analyze here. 

In the case of a one-component system, the simple shear flow has been 
extensively studied. Thus, Lun {\em et al.}\cite{LSJC84} obtained the 
rheological properties of a dense gas for small inelasticity, while Jenkins 
and Richman\cite{JR88} used a maximum-entropy approximation to solve the 
Enskog equation. An extension of the Jenkins and Richman work 
\cite{JR88} to highly inelastic spheres has been recently 
made\cite{CR98,C01}. For low-density granular gases, Sela {\em et 
al.}\cite{SGN96} have been able to get a perturbation solution of the 
Boltzmann equation to third order in the shear rate, finding normal stress 
differences at this level of approximation. On the other hand, some 
progresses have been done by using model kinetic equations in the 
low-density limit \cite{BMM97} as well as for dense gases \cite{MGSB99}. In 
both works, comparison with Monte Carlo simulations shows an excellent 
agreement even for strong dissipation.
Similar studies for multicomponent systems are more scarce.
Most of them are based on a Navier-Stokes description of the hydrodynamic
fields\cite{JM89,Z95,AW98,WA99} and, therefore, they are restricted to small 
shear rates, which for the steady shear flow is equivalent to the 
low-dissipation limit. As said before, although these studies permit
different temperatures for the two species, they lead to equal partial
granular temperatures $T_i$ in the quasielastic limit.
A primary attempt to include temperature differences was made by
Jenkins and Mancini \cite{JM87}, although applications of this theory which
appear in the literature incorporate the assumption of equipartition of
energy\cite{note}. Since this assumption is not
completely justified \cite{GD99,MP99,BRGD99,HWRL00,SD01}
for highly inelastic spheres, the problem of describing
the simple shear flow from a {\em multi-temperature} theory is still open.

The aim of this paper is to get the rheological properties of a binary granular 
mixture subjected to the simple shear flow in the framework of the Boltzmann 
equation. Two complementary routes are followed. First, the set of 
coupled Boltzmann equations are solved by using a first-Sonine 
polynomial approximation with a Gaussian measure. The main 
characteristic of our solution is that the reference Gaussian 
distributions are
defined in terms of the kinetic temperatures $T_i$ instead of the 
mixture temperature $T$. Consequently,
we do not assume {\em a priori} the equality of the 
three temperatures and the temperature ratio $T_1/T_2$ is 
consistently determined from the solution to the Boltzmann equations.
It is found that the partial temperatures of each species
are clearly different and so, the energy
is not equally distributed between both species. The consequences of this 
effect on the rheological properties are significant, as shown below. Once 
the temperature ratio is known, we get explicit expressions for the elements 
of the pressure tensor. The results are general and no limited 
to weak inelasticity or specific values of the parameters of the mixture. 
As a second alternative and to test the reliability of the theoretical 
predictions, we have used the Direct Simulation Monte Carlo (DSMC) method 
\cite{B94} to numerically solve the Boltzmann equation in the simple shear 
flow. Although the DSMC method was originally devised for molecular fluids, 
its extension to dealwith inelastic collisions is easy \cite{F00,MG01}.
For the elements of
the pressure tensor the agreement between theory and simulation turns out to 
be very good over a wide range of values of the restitution coefficients, 
mass ratios, concentration ratios and size ratios. It must be 
noted that in this paper we are interested in analyzing transport 
properties in the {\em uniform} shear flow. As several authors have shown, 
\cite{varios} the simple shear flow is unstable to long enough
wavelength perturbations so that clusters of particles are spontaneously
developed throughout the system. Here, we will restrict ourselves to the
{\em uniform} case, assuming that the system has reached such a state, and
without paying attention to the possible formation of particle clusters
(microstructure).

The plan of the paper is as follows. In Sec.\ \ref{sec2}, the coupled
Boltzmann equations and the corresponding hydrodynamic equations are 
recalled. The steady shear flow problem is also introduced in Sec.\ 
\ref{sec2}, while the Sonine approximation is discussed in Sec.\ \ref{sec3}. 
Section \ref{sec4} deals with the Monte Carlo simulation of the Boltzmann 
equation particularized for steady simple shear flow. The 
comparison between theory and simulation is presented in Sec.\ \ref{sec5} 
and we close the paper in Sec.\ \ref{sec6} with a short discussion.

\section{The Boltzmann equation and the simple shear flow}
\label{sec2}

We consider a binary mixture of smooth hard spheres of masses $m_{1}$ and 
$m_{2}$ and diameters $\sigma _{1}$ and $\sigma _{2}$. The 
inelasticity of collisions are characterized by three independent constant 
coefficients of normal restitution $\alpha_{11}$, $\alpha_{22}$, and 
$\alpha_{12}=\alpha_{21}$, where $\alpha_{ij}$ is the restitution 
coefficient for collisions between particles of species $i$ with $j$. In the
low-density regime, the distribution functions $f_{i}({\bf r},{\bf v};t)$ $
(i=1,2)$ for the two species verify the set of nonlinear Boltzmann equations 
\cite{GD99}  
\begin{equation}
\left( \partial _{t}+{\bf v}_{1}\cdot \nabla \right) f_{i}({\bf r},{\bf v}
_{1},t)=\sum_{j}J_{ij}\left[ {\bf v}_{1}|f_{i}(t),f_{j}(t)\right] \;.
\label{2.1}
\end{equation}
The Boltzmann collision operator $J_{ij}\left[ {\bf v}_{1}|f_{i},f_{j}\right]
$ describing the scattering of pairs of particles is 
\begin{eqnarray}
J_{ij}\left[{\bf v}_{1}|f_{i},f_{j}\right]  &=&\sigma _{ij}^{2}\int d{\bf v}
_{2}\int d\widehat{\bbox {\sigma }}\,\Theta (\widehat{\bbox {\sigma}}\cdot 
{\bf g}_{12})(\widehat{\bbox {\sigma }}\cdot {\bf g}_{12})  \nonumber \\
&&\times \left[ \alpha_{ij}^{-2}f_{i}({\bf r},{\bf v}_{1}^{\prime},t)f_{j}(
{\bf r},{\bf v}_{2}^{\prime },t)-f_{i}({\bf r},{\bf v}_{1},t)f_{j}({\bf r},
{\bf v}_{2},t)\right] \;,  \label{2.2}
\end{eqnarray}
where $\sigma_{ij}=\left( \sigma_{i}+\sigma_{j}\right)/2$, $\widehat{
\bbox {\sigma}}$ is a unit vector along their line of centers, $\Theta $ is
the Heaviside step function, and ${\bf g}_{12}={\bf v}_{1}-{\bf v}_{2}$. The
primes on the velocities denote the initial values $\{{\bf v}_{1}^{\prime},
{\bf v}_{2}^{\prime}\}$ that lead to $\{{\bf v}_{1},{\bf v}_{2}\}$
following a binary collision: 
\begin{equation}
{\bf v}_{1}^{\prime }={\bf v}_{1}-\mu _{ji}\left( 1+\alpha_{ij}    
^{-1}\right)(\widehat{\bbox {\sigma}}\cdot {\bf g}_{12})\widehat{\bbox 
{\sigma}},
\quad {\bf v}_{2}^{\prime}={\bf v}_{2}+\mu_{ij}\left( 
1+\alpha_{ij}^{-1}\right) (\widehat{\bbox {\sigma}}\cdot {\bf 
g}_{12})\widehat{\bbox{\sigma}}\;,  \label{2.3}
\end{equation}
where $\mu_{ij}=m_{i}/\left( m_{i}+m_{j}\right) $. The relevant
hydrodynamic fields are the number densities $n_{i}$, the flow velocity $
{\bf u}$, and the temperature $T$. They are defined in terms of moments of
the distributions $f_{i}$ as 
\begin{equation}
n_{i}=\int d{\bf v}_{1}f_{i}({\bf v}_{1})\;,\quad \rho {\bf u}
=\sum_{i}\rho_i{\bf u}_i=\sum_{i}\int d{\bf v}_{1}m_{i}{\bf v}_{1}f_{i}({\bf 
v}_{1})\;, 
 \label{2.4}
\end{equation}
\begin{equation}
nT=\sum_in_iT_i=\sum_{i}\int d{\bf v}_{1}\frac{m_{i}}{3}
V_1^{2}f_{i}({\bf v}_{1})\;,  \label{2.5}
\end{equation}
where $n=n_{1}+n_{2}$ is the total number density, $\rho 
=\rho_1+\rho_2=m_{1}n_{1}+m_{2}n_{2}$ is the total mass density, and ${\bf 
V}_1={\bf v}_1-{\bf u}$ is the peculiar velocity. Equations (\ref{2.4}) and 
(\ref{2.5}) also define the flow velocity ${\bf u}_i$ and the partial 
temperature $T_i$ of species $i$.  

The collision operators conserve the number of particles of each species 
and the total momentum, but the total energy is not conserved: 
\begin{equation}
\int d{\bf v}_{1}J_{ij}[{\bf v}_{1}|f_{i},f_{j}]=0\;,  \label{2.6}
\end{equation}
\begin{equation}
\sum_{i,j}\int d{\bf v}_{1}m_{i}{\bf v}_{1}J_{ij}[{\bf v}_{1}|f_{i},f_{j}]=0
\;,  \label{2.7}
\end{equation}
\begin{equation}
\sum_{i,j}\int d{\bf v}_{1}\case{1}{2}m_{i}V_{1}^{2}J_{ij}[{\bf v}
_{1}|f_{i},f_{j}]=-\case{3}{2}nT\zeta \;,  
\label{2.8}
\end{equation}
where $\zeta$ is identified as the ``cooling rate'' due to inelastic
collisions among all species. At a kinetic level, it is also convenient to 
discuss energy transfer in terms of the ``cooling rates'' $\zeta_i$ for the 
partial temperatures $T_i$. They are defined as 
\begin{equation}
\label{2.8.1}
\zeta_i=-\frac{2}{3n_iT_i}\sum_j\int 
d{\bf v}_{1}\case{1}{2}m_{i}V_{1}^{2}J_{ij}[{\bf 
v}_{1}|f_{i},f_{j}]\;.
\end{equation}
The total cooling rate $\zeta$ can be written as 
\begin{equation} 
\label{2.8.2}
\zeta=T^{-1}\sum_i\ x_iT_i\zeta_i\;,
\end{equation}
$x_i=n_i/n$ being the molar fraction of species $i$.

{}From Eqs.\ (\ref{2.4})--(\ref{2.8}), the macroscopic balance equations for 
the mixture can be obtained. They are given by 
\begin{equation}
D_{t}n_{i}+n_{i}\nabla \cdot {\bf u}+\frac
{\nabla \cdot {\bf j}_{i}}{m_{i}}=0\;,  \label{2.9}
\end{equation}
\begin{equation}
D_{t}{\bf u}+\rho ^{-1}\nabla \cdot {\sf P}=0\;,  \label{2.10}
\end{equation}
\begin{equation}
D_{t}T-\frac{T}{n}\sum_{i}\frac{\nabla \cdot {\bf j}_{i}}{m_{i}}+\frac{2}{
3n}\left( \nabla \cdot {\bf q}+{\sf P}:\nabla {\bf u}\right) =-\zeta
\,T\;.  \label{2.11}
\end{equation}
In the above equations, $D_{t}=\partial _{t}+{\bf u}\cdot \nabla $ is the
material derivative, 
\begin{equation}
{\bf j}_{i}=m_{i}\int d{\bf v}_1\,{\bf V}_1\,f_{i}({\bf V}_1)
\label{2.11bb}
\end{equation}
is the mass flux for species $i$ relative to the local flow, 
\begin{equation}
{\sf P}=\sum_i {\sf P}_i=\sum_{i}\,\int d{\bf v}_1\,m_{i}{\bf V}_1{\bf 
V}_1\,f_{i}({\bf V}_1) 
\label{2.12}
\end{equation}
is the total pressure tensor, and  
\begin{equation}
{\bf q}=\sum_i {\bf q}_i=\sum_{i}\,\int d{\bf 
v}_1\,\case{1}{2}m_{i}V_1^{2}{\bf V}_1\,f_{i}({\bf V}_1)
\label{2.13}
\end{equation}
is the total heat flux. The partial contributions to the pressure tensor, 
${\sf P}_i$, and the heat flux, ${\bf q}_i$, coming from species $i$ can be 
identified from Eqs.\ (\ref{2.12}) and (\ref{2.13}). 

As said in the Introduction, here we are interested in evaluating the 
rheological properties of a granular binary mixture subjected to the simple 
shear flow. From a macroscopic point of view, this state is characterized by 
a constant linear velocity profile ${\bf u}={\bf u}_i={\sf a}\cdot {\bf r}$, 
where the elements of the tensor ${\sf a}$ are 
$a_{k\ell}=a\delta_{kx}\delta_{\ell y}$, $a$ being the constant shear rate. 
In addition, the partial densities $n_i$ and the granular temperature $T$
are uniform, while the mass and heat fluxes vanish by symmetry reasons. 
Thus, the (uniform) pressure tensor is the only nonzero flux in the problem. 
On the other hand, the temporal variation of the granular temperature arises 
from the balance of two opposite effects: viscous heating and dissipation in 
collisions. In the steady state both mechanisms cancel each other and the 
temperature remains constant. In that case, according to the balance energy 
equation (\ref{2.11}), the shear stress $P_{xy}$ and the cooling rate 
$\zeta$ are related by
\begin{equation}
\label{2.14}
aP_{xy}=-\frac{3}{2}\zeta p,
\end{equation}
where $p=nT$ is the pressure. Our aim is to analyze this 
steady state by means of an (approximate) analytical method as well as by  
performing Monte Carlo simulations of the Boltzmann equation. 

The simple shear flow becomes spatially uniform when one refers the 
velocities of the particles to a frame moving with the flow velocity ${\bf 
u}$: $f_i\left({\bf r},{\bf v}_1\right)\rightarrow f_i({\bf V}_1)$. 
Consequently, the corresponding Boltzmann equations (\ref{2.1}) read  
\begin{equation}
\label{2.15}
-a V_{1,y}\frac{\partial}{\partial 
V_{1,x}}f_1({\bf V}_1)=J_{11}[{\bf V}_1|f_1,f_1]+J_{12}[{\bf V}_1|f_1,f_2]\;,
\end{equation}
\begin{equation}
\label{2.16}
-a V_{1,y}\frac{\partial}{\partial 
V_{1,x}}f_2({\bf V}_1)=J_{22}[{\bf V}_1|f_2,f_2]+J_{21}[{\bf V}_1|f_2,f_1]\;.
\end{equation}
The elements of the partial pressure tensors ${\sf P}_i$ ($i=1,2$) can be 
obtained by multiplying the Boltzmann equations by $V_{1,k}V_{1,\ell}$ and 
integrating over ${\bf V}_1$. The result is
\begin{equation}
\label{2.17}
a_{k m}P_{1,\ell m}+a_{\ell m}P_{1,k m}=A_{k \ell}^{11}+A_{k\ell}^{12}
\quad (1\leftrightarrow2)\;,
\end{equation}
where 
\begin{equation}
\label{2.18}
A_{k\ell}^{ij}=m_i\int d{\bf V}_1 V_{1,k}V_{1,\ell} J_{ij}[{\bf 
V}_1|f_i,f_j]\;.
\end{equation}
{}From Eq.\ (\ref{2.17}), in particular, one obtains
\begin{equation}
\label{2.19}
a P_{1,xy}=-\case{3}{2}p_1\zeta_1\;,
\end{equation}
\begin{equation}
\label{2.20}
a P_{1,yy}=A_{xy}^{11}+A_{xy}^{12}\;,
\end{equation}
\begin{equation}
\label{2.21}
0=A_{yy}^{11}+A_{yy}^{12}=A_{zz}^{11}+A_{zz}^{12}\;.
\end{equation}
Here, $p_1=n_1T_1=(P_{1,xx}+P_{1,yy}+P_{1,zz})/3$ is the partial pressure of 
species $1$ and upon writing Eq.\ (\ref{2.19}) we have considered the 
relation (\ref{2.8}). The corresponding equations for ${\sf P}_2$ can be 
easily written just by interchanging the indices $1$ and $2$. Thus, the 
determination of the elements of the partial pressure tensors ${\sf P}_i$ is 
a closed problem once the cooling rates $\zeta_i$ and the collisional 
moments $A_{k\ell}^{ij}$ are known. This requires the explicit knowledge of 
the velocity distribution functions $f_i$. 

\section{Approximate solution}
\label{sec3}

Unfortunately, solving the Boltzmann equations (\ref{2.15}) and 
(\ref{2.16}) is a formidable task and it does not seem possible to get the 
exact forms of the distributions $f_i$, even in the one-component case. A 
possible way to overcome such a problem is to expand $f_i$ in a complete set 
of polynomials with a Gaussian measure and then truncate the series. In 
practice, Sonine polynomials are used. This approach is similar to the usual 
moment method for solving kinetic equations in the elastic case. In the 
context of granular gases, this strategy has been widely applied in the past 
few years in the one-component case as well as for multicomponent systems 
and excellent approximations have been obtained by retaining only the first 
two terms. Therefore, one can expect to get a reasonable estimate for 
$\zeta_i$ and $A_{k\ell}^{ij}$ by using the following approximation for 
$f_i$: 
\begin{equation}
\label{2.22}
f_i({\bf V}_1)\to 
f_{i,M}({\bf V}_1)\left[1+\frac{m_i}{2T_i}C_{i,k\ell}\left(V_{1,k}V_{1,\ell}
-\frac{1}{3}V_1^2\delta_{k\ell}\right)\right],
\end{equation}
where $f_{i,M}$ is a Maxwellian distribution at the temperature of the 
species $i$, i.e.,
\begin{equation}
\label{2.23} 
f_{i,M}({\bf V}_1)=n_i 
\left(\frac{m_i}{2\pi 
T_i}\right)^{3/2}\exp\left(-\frac{m_iV^2}{2T_i}\right). 
\end{equation} 
As we will show later, in general the three temperatures 
$T$, $T_1$, and $T_2$ are different in the inelastic case. For this reason
we choose the parameters in the Maxwellians so that it is  
normalized to $n_i$ and provides the exact second moment of $f_i$.
The Maxwellians $f_{i,M}$ for the two species
can be quite different due to the temperature differences. This aspect is
essential in our two-temperature theory and has not been taken into account
in all previous studies. The coefficient ${\sf C}_i$ can be identified by requiring the moments 
with respect to $V_{1,k}V_{1,\ell}$ of the trial function (\ref{2.22}) to be the same as those for the exact 
distribution $f_i$. This leads to 
\begin{equation}
\label{2.24}
C_{i,k\ell}=\frac{P_{i,k\ell}}{p_i}-\delta_{k\ell}.
\end{equation}
With this approximation, the integrals appearing in the expressions of 
$\zeta_i$ and $A_{k\ell}^{ij}$ can be evaluated and the details are given in 
Appendices \ref{appA} and \ref{appB}.

In order to express the solution of the system of equations for the pressure 
tensor, it is convenient to introduce dimensionless quantities. Thus, we 
introduce the reduced cooling rates $\zeta_i^*=\zeta_i/\nu$, 
the reduced temperature $T^*=\nu^2/a^2$, and the reduced 
pressure tensors ${\sf P}_i^*={\sf P}_i/x_i p$. The (reduced) total pressure 
tensor ${\sf P}^*={\sf P}/p=x_1{\sf P}_{1}^*+x_2{\sf P}_2^*$.  Here, 
$\nu=\sqrt{\pi}n\sigma_{12}^2v_0$ is a characteristic collision frequency 
and $v_0=\sqrt{2T(m_1+m_2)/m_1m_2}$ is a thermal velocity defined in terms 
of the temperature of the mixture $T$. Notice that, for given values of the 
parameters of the mixture, $T^{*}$ and ${\sf P}^*$ are  
functions of the restitution coefficients $\alpha_{ij}$ only. 

According to the symmetry of the problem, $P_{i,xz}=P_{i,yz}=0$, so that the 
nonzero elements are $P_{i,xx}$, $P_{i,yy}$, $P_{i,zz}$, and 
$P_{i,xy}=P_{i,yx}$. The three normal elements are not independent since 
$P_{i,xx}^*+P_{i,yy}^*+P_{i,zz}^*=3\gamma_i$, where the temperature ratios 
$\gamma_i=T_i/T$ are given by 
\begin{equation}
\label{2.25}
\gamma_1=\frac{\gamma}{1+x_1(\gamma-1)},\quad 
\gamma_2=\frac{1}{1+x_1(\gamma-1)}\;,
\end{equation}
with $\gamma=T_1/T_2$. The temperature ratio $\gamma$ provides information 
about how the kinetic energy is distributed between both species. In 
addition, according to Eq.\ (\ref{2.21}), $P_{i,yy}^*=P_{i,zz}^*$. 
Consequently, the partial pressure tensors have four relevant elements, say 
for instance: ${\sf {\cal P}}\equiv \{P_{1,yy}^*, P_{2,yy}^*, P_{1,xy}^*, 
P_{2,xy}^{*}\}$. Taking into account the results derived in the Appendices, 
Eqs.\ (\ref{2.20}) and (\ref{2.21}) (plus their corresponding counterparts 
for species $2$) can be written as
\begin{equation}
\label{2.26} 
{\sf {\cal L}}{\sf{\cal P}}={\sf {\cal Q}},
\end{equation}
where ${\sf {\cal L}}$ is the $4\times 4$ matrix 
\begin{equation}
\label{2.26.1}
{\sf {\cal L}}=\left(
\begin{array}{cccc}
1&0&-(G_{11}+G_{12})\nu/a&-H_{12}\nu/a\\
0&1&-H_{21}\nu/a&-(G_{22}+G_{21})\nu/a\\
-(G_{11}+G_{12})&-H_{12}&0&0\\
-H_{21}&-(G_{22}+G_{21})&0&0
\end{array}
\right),
\end{equation}
and 
\begin{equation}
\label{2.26.2}
{\sf {\cal Q}}=\left(
\begin{array}{c}
0\\
0\\
F_{11}+F_{12}\\
F_{22}+F_{21}
\end{array}
\right).
\end{equation}
Here, the functions $F_{ij}$, $G_{ij}$, and $H_{ij}$ are defined 
in the Appendix \ref{appB}. The solution to Eq.\ (\ref{2.26}) is
\begin{equation}
\label{2.27}
{\sf {\cal P}}={\sf {\cal L}}^{-1}{\sf {\cal Q}}\;.
\end{equation}
Equation (\ref{2.27}) gives the nonzero elements of the pressure tensors 
${\sf P}_i^*$ in terms of the reduced temperature $T^*$ (or the 
reduced shear rate $a/\nu$), the temperature ratio $\gamma$, the 
restitution coefficients and the parameters of the mixture. The dependence 
of $T^*$ on the coefficients $\alpha_{ij}$ can be obtained from the energy 
balance equation (\ref{2.14})
\begin{equation}
\label{2.28}
T^{*-1/2}=-\frac{3}{2}\frac{\zeta^*}{P_{xy}^*}=
-\frac{3}{2}\frac{x_1\gamma_1\zeta_1^*+x_2\gamma_2\zeta_2^*}{x_1P_{1,xy}^*+
x_2P_{2,xy}^*}.
\end{equation}
Finally, when Eqs.\ (\ref{2.27}) and (\ref{2.28}) are used in Eq.\ 
(\ref{2.19}) (or its counterpart for the species 2), one gets a ${\em 
closed}$ equation for the temperature ratio $\gamma$, that can be solved 
numerically. In reduced units, this equation can be written as 
\begin{equation}
\label{2.29}
\gamma=
\frac{\zeta_2^*}{\zeta_1^*}\frac{P_{1,xy}^*}{P_{2,xy}^*}.
\end{equation}

In the elastic limit ($\alpha_{ij}=1$), we recover previous results 
derived for molecular gases\cite{MGS95}. A simple and interesting case 
corresponds to the case of mechanically equivalent particles ($m_1=m_2$, 
$\alpha_{11}=\alpha_{22}=\alpha_{12}\equiv\alpha$, 
$\sigma_{11}=\sigma_{22}$). In this limit, Eqs.\ (\ref{2.27})--(\ref{2.29}) 
leads to $\gamma=1$, ${\sf P}^*={\sf P}_{1}^*={\sf P}_{2}^*$, with   
\begin{equation}
\label{2.30}
P_{yy}^*=\frac{2}{3}\frac{2+\alpha}{3-\alpha}\;,
\end{equation}
\begin{equation}
\label{2.31}
P_{xy}^*=-\frac{5}{3}\frac{2+\alpha}{(1+\alpha)(3-\alpha)^2}\frac{a}{\nu}\;,
\end{equation}
\begin{equation}
\label{2.32}
P_{xx}^*=3-2P_{yy}^*,
\end{equation}
and 
\begin{equation}
\label{2.33}
T^{*-1}\equiv\frac{a^2}{\nu^2}=\frac{3}{5}\frac{(1+\alpha)(3-\alpha)^2}
{2+\alpha}(1-\alpha^2)\;.
\end{equation}
These expressions differ from the results derived in Ref.\ 
\onlinecite{BMM97} by using a model kinetic equation. However, for practical 
purposes, the discrepancies between both approximations are quite small, 
even for moderate values of the restitution coefficient. It is also 
interesting to consider the limit of weak dissipation
($1-\alpha_{ij}\ll 1$), in which case it is
possible to get analytical results. For the sake of simplicity, let us 
assume that all the particles have the same coefficient of restitution, 
namely, $\alpha_{11}=\alpha_{22}=\alpha_{12}\equiv \alpha$. To get the first 
order corrections in the quasielastic limit, we introduce the perturbation 
parameter $\epsilon\equiv (1-\alpha^2)^{1/2}$ and perform a series expansion 
around $\epsilon=0$. The details of the calculation are presented in 
Appendix \ref{appC} and here we only quote the final results. First, the 
leading term of the reduced shear rate $a/\nu$ (which is a measure of the 
steady granular temperature) is
\begin{equation}
\label{2.34}
\frac{a}{\nu}=a_0 (1-\alpha^2)^{1/2}+\cdots.
\end{equation}
Next, the temperature ratio and the relevant elements of the (partial) 
pressure tensors can be written as
\begin{equation}
\label{2.35}
\gamma=1+\gamma_0 a_0^2 (1-\alpha^2)+\cdots,
\end{equation}
\begin{equation}
\label{2.36}
P_{i,yy}^*=1+P_{i,yy}^{(2)}a_0^2 (1-\alpha^2)+\cdots,
\end{equation}
\begin{equation}
\label{2.37}
P_{i,xx}^*=1+P_{i,xx}^{(2)} a_0^2 (1-\alpha^2)+\cdots,
\end{equation}
\begin{equation}
\label{2.38}
P_{i,xy}^*=P_{i,xy}^{(1)}a_0 (1-\alpha^2)^{1/2}+\cdots.
\end{equation}
In these equations, $a_0$, $\gamma_0$, and $P_{i,k\ell}^{(r)}$ are 
dimensionless coefficients that depend on the ratios of mass, concentrations 
and sizes. Their explicit expressions are given in Appendix \ref{appC}.

In summary, by using the Sonine approximation (\ref{2.22}), we have 
explicitly determined the rheological properties of the mixture as well as 
the reduced shear rate and the temperature ratio as functions of 
dissipation and mechanical parameters of the mixture. These 
theoretical predictions will be compared with those obtained from Monte 
Carlo simulations in Sec.\ \ref{sec5}.

\section{Monte Carlo simulation}
\label{sec4}

{}From a numerical point of view, the Direct Simulation Monte Carlo (DSMC) 
method\cite{B94} is the most convenient algorithm to study non-equilibrium 
phenomena in the low-density regime. It was devised to mimic the dynamics 
involved in the Boltzmann collision term. 
The extension of the DSMC method to deal with inelastic collisions is 
straightforward\cite{MGSB99,F00,MG01}, and here we have used it to 
numerically solve the Boltzmann equation in the simple shear flow. In 
addition, since the simple shear flow is spatially homogeneous in the local 
Lagrangian frame, the simulation method becomes especially easy to 
implement. This is an important advantage with respect to molecular dynamics 
simulations. On the other hand, the restriction to this homogeneous state 
prevents us from analyzing the possible instability of simple shear flow or 
the formation of clusters or microstructures. 

The DSMC method as applied to the simple shear flow is as follows. 
The velocity distribution function of the species $i$ is represented by the 
peculiar velocities $\{{\bf V}_k\}$ of $N_i$ ``simulated" particles:
\begin{equation}
\label{4.1}
f_i({\bf V},t)\to n_i \frac{1}{N_i}\sum_{k=1}^{N_i} \delta({\bf V}-{\bf 
V}_k(t))\; .
\end{equation}
Note that the number of particles $N_i$ must be taken according to the 
relation $N_1/N_2=n_1/n_2$. At the initial state, one assigns velocities to 
the particles drawn from the Maxwell-Boltzmann probability distribution: 
\begin{equation}
\label{4.2}
f_i({\bf V},0)=n_i\ \pi^{-3/2}\ V_{0i}^{-3}(0)\ 
\exp\left(-V^2/V_{0i}^2(0)\right)\;,
\end{equation}
where $V_{0i}^2(0)=2T(0)/m_i$ and $T(0)$ is the initial temperature. To 
enforce a vanishing initial total momentum, the velocity of every particle 
is subsequently subtracted by the amount $N_i^{-1} \sum_k {\bf V}_k(0)$. In 
the DSMC method, the free motion and the collisions are uncoupled over a 
time step $\Delta t$ which is small compared with the mean free time and the 
inverse shear rate. In the local Lagrangian frame, particles of each species 
($i=1,2$) are subjected to the action of a non-conservative inertial force 
${\bf F}_i=-m_i\ {\sf a}\cdot{\bf V}$. This force is represented by the 
terms on the left-hand side of Eqs.\ (\ref{2.15}) and (\ref{2.16}). Thus, 
the free motion stage consists of making ${\bf V}_k\to {\bf V}_k-{\sf 
a}\cdot{\bf V}_k\Delta t$. In the collision stage, binary interactions 
between particles of species $i$ and $j$ must be considered. To simulate the 
collisions between particles of species $i$ with $j$ a sample of 
$\frac{1}{2} N_i \omega_{\text{max}}^{(ij)}\Delta t$ pairs is chosen at 
random with equiprobability. Here, $\omega_{\text{max}}^{(ij)}$ is an upper 
bound estimate of the probability that a particle of the species $i$ 
collides with a particle of the species $j$. 
Let us consider a pair $\{k,\ell\}$ belonging to this sample. Here, $k$ 
denotes a particle of species $i$ and $\ell$ a particle of species $j$. 
For each pair $\{k,\ell\}$ with velocities 
$\{{\bf V}_k,{\bf V}_{\ell}\}$, the following steps 
are taken: (1) a given direction $\widehat{\bbox \sigma}_{k\ell}$ is chosen 
at random with equiprobability; (2) the collision between particles $k$ and 
$\ell$ is accepted with a probability equal to $\Theta({\bf g}_{k\ell}\cdot 
\widehat{\bbox \sigma}_{k\ell})\omega_{k\ell}^{(ij)}/
\omega_{\text{max}}^{(ij)}$, where 
$\omega_{k\ell}^{(ij)}=4\pi \sigma_{ij}^2 n_j|{\bf g}_{k\ell}\cdot 
\widehat{\bbox \sigma}_{k\ell}|$ and ${\bf g}_{k\ell}={\bf V}_k-{\bf 
V}_{\ell}$; (3) if the collision is accepted, postcollisional velocities are 
assigned to both particles according to the scattering rules:
\begin{equation}
\label{4.3}
{\bf V}_{k}\to {\bf V}_{k}-\mu_{ji}
(1+\alpha_{ij})({\bf g}_{k\ell}\cdot \widehat{\bbox 
\sigma}_{k\ell})\widehat{\bbox \sigma}_{k\ell}\; ,
\end{equation}
\begin{equation}
\label{4.4}
{\bf V}_{\ell}\to {\bf V}_{\ell}+\mu_{ij}
(1+\alpha_{ij})({\bf g}_{k\ell}\cdot \widehat{\bbox 
\sigma}_{k\ell})\widehat{\bbox \sigma}_{k\ell}\;.
\end{equation}
In the case that in one of the collisions 
$\omega_{k\ell}^{(ij)}>\omega_{\text{max}}^{(ij)}$, the estimate of 
$\omega_{\text{max}}^{(ij)}$ is updated as 
$\omega_{\text{max}}^{(ij)}=\omega_{k\ell}^{(ij)}$. The procedure described 
above is performed for $i=1,2$ and $j=1,2$.

In the course of the simulations, one evaluates the total pressure tensor 
and the partial temperatures. They are given as 
\begin{equation}
\label{4.5}
{\sf P}=\sum_{i=1}^{2} \frac{m_i n_i}{N_i}\sum_{k=1}^{N_i} {\bf V}_k {\bf 
V}_k\; ,
\end{equation}
\begin{equation}
T_i=\frac{m_i}{3N_i}\sum_{k=1}^{N_i}{\bf V}_k^2\; .
\end{equation}
To improve the statistics, the results are averaged over a number 
${\cal N}$ of independent realizations or replicas. In our simulations we 
have typically taken a total number of particles $N=N_1+N_2=10^5$, a number 
of replicas ${\cal N}=10$, and a time step $\Delta t=3\times 10^{-3}
\lambda_{11}/V_{01}(0)$, where 
$\lambda_{11}=(\sqrt{2} \pi n_1 \sigma_{11}^2)^{-1}$ is the mean free path 
for collisions 1--1.

A complete presentation of the results is complex since there are many 
parameters involved: $\{\alpha_{ij}, m_1/m_2, n_1/n_2, 
\sigma_{11}/\sigma_{22}\}$. For the sake of concreteness, henceforth we will 
consider the case $\alpha_{11}=\alpha_{22}=\alpha_{12}\equiv\alpha$. 
In the steady state, the reduced quantities $T^*$, $\gamma_i$, and ${\sf 
P}^*$ are independent of the initial state for given 
values of the restitution coefficient and the ratios of mass, concentration 
and sizes. To illustrate it, in Fig.\ \ref{fig1} we present the time 
evolution of $T^*(t)$ for $\alpha=0.75$, 
$\sigma_{11}/\sigma_{22}=1$, $m_1/m_2=4$, $n_1/n_2=1/3$ and three different 
initial conditions. Time is measured in units of $\lambda_{11}/V_{01}(0)$. 
After an initial transient period, all curves converge to the same steady 
value, as predicted by the solution described in Sec.\ \ref{sec3}. The same 
qualitative behavior has been found for the temperature ratio and the 
elements of the reduced pressure tensor. Therefore, in the following we will 
focus on the dependence of the steady values of the reduced quantities on 
the restitution coefficient $\alpha$ and the parameters of the mixture, once 
we have checked they do not depend on the initial state.

\section{Comparison between theory and Monte Carlo simulations}
\label{sec5} 
 
In this Section we compare the predictions of the Sonine approximation with 
the results obtained from the DSMC method. Our goal is to explore the 
dependence of $a^*$, $\gamma=T_1/T_2$ and the nonzero elements of ${\sf 
P}^*$ on $\alpha$, the mass ratio $\mu\equiv m_1/m_2$, the 
concentration ratio $\delta\equiv n_1/n_2$, and the ratio of sizes 
$w\equiv \sigma_{11}/\sigma_{22}$. 

First, we will investigate the dependence of the relevant quantities on 
$\alpha$ and $\mu$ for given values of $\delta$ and $w$. Recent molecular 
dynamics simulations for a dilute monocomponent system of smooth inelastic 
hard disks\cite{GT96} have supported an ``equation of state'' to a
sheared granular system in which the steady (reduced) temperature $T^*$ can 
be closely fitted by a linear function of $(1-\alpha^2)^{-1}$. Similar 
results have been obtained from kinetic models of the Boltzmann \cite{BMM97} 
and Enskog \cite{MGSB99} equations. An interesting question is whether this 
simple relationship can be extended to the case of multicomponent systems. 
The results obtained here (both from the simulations and from the kinetic 
theory analysis) for mixtures of different masses, concentrations or sizes 
show that $T^{*}$ is indeed a {\em quasi}-linear function of 
$(1-\alpha^2)^{-1}$. As an illustrative example, we consider the case $w=1$, 
$\delta=1$ (equimolar mixture), and three different values of the mass ratio 
$\mu=1$, $2$, and $10$. Figure \ref{fig2} shows $T^{*}$ versus 
$(1-\alpha^2)^{-1}$ as obtained from the simulations (symbols) and from the 
Sonine approximation (lines). It is evident that the kinetic theory has an 
excellent agreement with the simulation results and also that $T^{*}$ is 
practically linear in $(1-\alpha^2)^{-1}$ whatever the mass ratio considered 
is. The slope of the straight lines increases as the disparity of masses 
increases. 

The temperature ratio and the non-zero elements of the pressure tensor are 
plotted in Figs.\ \ref{fig3} and \ref{fig4} $(a-d)$, respectively, as a 
function of the dissipation parameter $\alpha$ for the same cases as those 
considered in Fig.\ \ref{fig2}. The curves corresponding to $\mu<1$ can be 
easily inferred from them. Figure \ref{fig3} clearly shows that, except 
for mechanically equivalent particles, the partial temperatures are 
different, even for moderate dissipation (say $\alpha\simeq 0.9$). This 
means that the traditional assumption of equipartition of fluctuation energy
begins to fail.
This effect is generic of multicomponent dissipative systems and is
consistent with results recently derived in the homogeneous cooling 
state\cite{GD99,SD01} as well as in driven systems\cite{FM02,CH02}. 
The extent of the equipartition violation depends
on the concentrations and the mechanical differences of the particles (e.g.,
masses, sizes, restitution coefficients), and is greater when the differences
are large. The agreement between theory and simulation is again excellent.
With respect to the pressure tensor, Fig.\ \ref{fig4} $(a-d)$, we see that 
the theory captures well the main trends observed for the rheological 
properties. At a quantitative level, the agreement is better in the case of 
the shear stress $P_{xy}^*$ and the normal element $P_{xx}^*$, while the 
discrepancies for the normal stresses $P_{yy}^*$ and $P_{zz}^*$ are larger 
than in the case of the temperature ratio, especially as the restitution 
coefficient decreases. On the other hand, the theory only predicts normal 
stress differences in the plane of shear flow ($P_{xx}^*\neq 
P_{yy}^*=P_{zz}^*$) while the simulation also shows that there is anisotropy 
in the plane perpendicular to the flow velocity, $P_{zz}^*> P_{yy}^*$. This 
kind of anisotropy has been also observed in molecular dynamics simulations 
of shear flows \cite{HS92}. Nevertheless, these relative normal stress 
differences in this plane are very small and decrease as $\alpha$ increases.

The influence of the concentration ratio $\delta$ on the temperature ratio 
and the rheological properties is shown in Figs.\ \ref{fig5} and \ref{fig6} 
$(a-d)$, respectively, for $w=1$, $\mu=4$, and two values of $\delta$: 
$\delta=1/3$ and $\delta=3$. We observe again a strong dependence of the 
temperature ratio $\gamma$ on dissipation. For a given value of $\alpha$, 
the temperature ratio increases as the molar fraction of the heavy species 
decreases. Concerning shear stresses, we see that they are practically 
independent of the concentration ratio since all the curves collapse in a 
common curve. A more significant influence is observed for the normal 
stresses. In general, the agreement with the theory is good although the 
discrepancies are more important in the case of $\delta=1/3$. 
Finally, the influence of the size of the particles on the rheological 
properties is illustrated in Fig.\ \ref{fig7} $(a-d)$. We consider 
an equimolar mixture ($\delta=1$) of particles of equal mass ($\mu=1$) for 
two different values of the size ratio: $w=1$, and $w=2$. We see that 
similar conclusions to those previously found in Figs.\ 
\ref{fig3}--\ref{fig6} are obtained when one considers mixtures of particles 
of different sizes.

\section{Discussion}
\label{sec6}

In this paper we have addressed the problem of a low-density granular 
mixture constituted by smooth inelastic hard spheres and subjected to a 
linear shear flow $u_x=ay$. We are interested in the steady state where the 
effect of viscosity is compensated for by the dissipation in collisions. Our 
description applies for arbitrary values of the shear rate $a$ or the 
inelasticity of the system and not restriction on the values of masses, 
concentrations and sizes are imposed in the system. The study has 
been made by using two different and complementary routes. On the one hand, 
the set of coupled Boltzmann equations are solved by means a Sonine 
polynomial approximation  and, on the other hand, Monte Carlo simulations 
are performed to numerically solve the Boltzmann equations.
Given that the partial temperatures $T_i$ of each species can be different,
the reference Maxwellians in the Sonine expansion are defined at the
temperature for that species. This is one of the new features of our
expansion. On the other hand, to put this work
in a proper context, it must be noticed that we have restricted our 
considerations to states in which the only gradient is the one associated 
with the shear rate so that density and velocity fluctuations are not 
allowed in the numerical simulation. 

We have focused on the analysis of the dependence of the steady 
(reduced) temperature $T^*$ and the (reduced) pressure tensor ${\sf P}^*$ on 
the coefficients of restitution $\alpha_{ij}$ and the parameters of the 
mixture, namely the mass ratio $\mu$, the concentration ratio $\delta$ and 
the size ratio $w$. The results clearly indicate that the deviation of the 
above quantities from their functional forms for elastic collisions is quite 
important, even for moderate dissipation. In particular, the temperature 
ratio, which measures the distribution of kinetic energy between both 
species, is different from unity and presents a complex dependence on the 
parameters of the problem. This result contrasts with previous results 
derived for granular mixtures\cite{JM89,Z95,AW98,WA99} where it was 
consistently assumed the equality of the partial temperatures in the small
inelasticity limit. In the same way
as in the homogeneous cooling state problem\cite{GD99,MG01},
the deviations from
the energy equipartition can be weak or strong depending on the mechanical 
differences between the species and the degree of dissipation. On the other
hand, the simulation as well as the theoretical results also show that 
the steady total temperature $T^*$ can be fitted 
by a linear function of $(1-\alpha^2)^{-1}$ with independence of the values 
of the parameters of the mixture. This extends previous results derived in 
the context of simple granular gases by using molecular dynamics \cite{GT96} 
or Monte Carlo simulations\cite{BMM97,MGSB99}. With respect to the 
rheological properties, comparison between theory and simulation shows a 
good quantitative agreement, especially for the shear stress $P_{xy}^*$, 
which is the most relevant element of the pressure tensor in a shearing
problem.  Although the kinetic theory also predicts normal stresses, the 
discrepancies between theory and simulation are larger than those found for 
the temperature ratio or the shear stress.

It is illustrative to make some comparison between the predictions
made from our two-temperature theory with those obtained if the
differences in the partial temperatures were neglected ($T_1=T_2=T$). For
instance, let us consider the mixture $\sigma_{11}=\sigma_{22}$, $n_1=n_2$,
and $m_1=10m_2$ with $\alpha=0.75$. In this case, for the $xy$ and $yy$ elements
of the pressure tensor, the simulation results are $-P_{xy}^*=0.498$ and
$P_{yy}^*=0.723$. Our two-temperature theory predicts $-P_{xy}^*=0.498$ and
$P_{yy}^*=0.743$ while the single-temperature theory (assumption made in
previous works) gives $-P_{xy}^*=0.456$ and $P_{yy}^*=0.815$. Clearly,
inclusion of the two-temperature effects improves the theoretical
estimations and makes a significant (quantitative) difference
with respect to the predictions of the single-temperature theory.

As a final comment, let us mention that the study made here can in principle be
extended in both aspects, kinetic theory and simulations, to the revised 
Enskog equation in order to assess the influence of finite density on the 
rheological properties of the mixture. Work along this line is in progress.  
 
\acknowledgments

Partial support from the Ministerio de Ciencia y Tecnolog\'{\i}a (Spain)
through Grant No. BFM 2001-0718 is acknowledged.

\appendix
\section{Evaluation of the cooling rates}
\label{appA}

In this Appendix the (reduced) cooling rates $\zeta_i^*$ are evaluated by 
using the first Sonine approximation (\ref{2.22}). The cooling rate is given 
by
\begin{equation}
\label{a1}
\zeta_i^*=-\frac{2}{3}\pi^{-1/2}\theta_i\sum_j\int d{\bf 
V}_1^*V_1^{*2}J_{ij}^*[{\bf V}_1^*|f_i^*,f_j^*],
\end{equation}
where $\theta_i=1/(\gamma_i\mu_{ji})$, ${\bf V}_1^*={\bf V}_1/v_0$, 
$J_{ij}^*=(v_0^2/n_in\sigma_{12}^2)J_{ij}$, and $f_i^*=(v_0^3/n_i)f_i$. 
Henceforth, it will be understood that dimensionless quantities will be used 
and the asterisks will be deleted to simplify the notation. A useful 
identity for an arbitrary function $h({\bf V}_{1})$ is given by 
\begin{eqnarray}
\int d{\bf V}_{1}h({\bf V}_{1})J_{ij}\left[ {\bf V}_{1}|f_{i},f_{j}\right]
&=&x_j\left(\frac{\sigma _{ij}}{\sigma_{12}}\right)^{2}\int \,d{\bf 
V}_{1}\,\int \,d{\bf V}_{2}f_{i}(
{\bf V}_{1})f_{j}({\bf V}_{2})  \nonumber \\
& &\times \int d\widehat{\bbox {\sigma}}\,\Theta (\widehat{\bbox {\sigma}}
\cdot {\bf g}_{12})(\widehat{\bbox {\sigma }}\cdot {\bf g}_{12})\,\left[ h(
{\bf V}_{1}^{^{\prime \prime }})-h({\bf V}_{1})\right] \;,  \label{a2}
\end{eqnarray}
with 
\begin{equation}
{\bf V}_{1}^{^{\prime \prime }}={\bf V}_{1}-\mu _{ji}(1+\alpha_{ij})(
\widehat{\bbox {\sigma }}\cdot {\bf g}_{12})\widehat{\bbox {\sigma}}\;.
\label{a3}
\end{equation}
This result applies for both $i=j$ and $i\neq j$. Use of this identity in 
Eq.\ (\ref{a1}) allows the angular integrals to be performed. The result is
\begin{eqnarray}
\label{a4}
\zeta_i&=&(1-\alpha_{ii}^2)\frac{1}{12}\sqrt{\pi} \theta_ix_i
\left(\frac{\sigma _{ii}}{\sigma_{12}}\right)^{2}\int \,d{\bf 
V}_{1}\,\int \,d{\bf V}_{2}\,g_{12}^3f_{i}({\bf V}_{1})f_{i}({\bf V}_{2}) 
\nonumber\\
& & +(1-\alpha_{ij}^2)\frac{1}{3}\sqrt{\pi} \theta_i\mu_{ji}^2x_j
\int \,d{\bf V}_{1}\,\int \,d{\bf V}_{2}\, g_{12}^3f_{i}({\bf 
V}_{1})f_{j}({\bf V}_{2}) \nonumber\\
& & +(1+\alpha_{ij})\frac{2}{3}\sqrt{\pi} \theta_i\mu_{ji}x_j
\int \,d{\bf V}_{1}\,\int \,d{\bf V}_{2}\,g_{12}
({\bf g}_{12}\cdot {\bf G}_{ij})f_{i}({\bf V}_{1})f_{j}({\bf V}_{2}) ,
\end{eqnarray}
where ${\bf G}_{ij}=\mu_{ij}{\bf V}_1+\mu_{ji}{\bf V}_2$. Now we consider 
the Sonine approximation (\ref{2.22}) for the distributions $f_i$:
\begin{equation}
\label{a4.1}
f_{i}({\bf V}_1)\to \left(\frac{\theta_i}{\pi}\right)^{3/2}e^{-\theta_i 
V_1^2}
\left[1+\theta_i C_{i,k\ell}\left(V_{1,k}V_{1,\ell}-\frac{1}{3}
V_1^2\delta_{k\ell}\right)\right].
\end{equation}
Neglecting nonlinear terms in the tensor $C_{i,k\ell}$, the expression 
(\ref{a4}) can be written as
\begin{eqnarray}
\label{a5}
\zeta_i&=&(1-\alpha_{ii}^2)\frac{1}{12}\pi^{-5/2} \theta_i^{-1/2}x_i
\left(\frac{\sigma _{ii}}{\sigma_{12}}\right)^{2}\int \,d{\bf 
V}_{1}\,\int \,d{\bf V}_{2}\,g_{12}^3\ e^{-(V_1^2+V_2^2)}
\nonumber\\
& & +(1-\alpha_{ij}^2)\frac{1}{3}\pi^{-5/2} 
(\theta_i\theta_j)^{3/2}\mu_{ji}^2x_j \theta_i
\int \,d{\bf V}_{1}\,\int \,d{\bf V}_{2}\,g_{12}^3
\ e^{-(\theta_iV_1^2+\theta_jV_2^2)}
\nonumber\\
& & +(1+\alpha_{ij})\frac{2}{3}\pi^{-5/2} (\theta_i\theta_j)^{3/2}
\mu_{ji}x_j\theta_i\int \,d{\bf V}_{1}\,\int \,d{\bf V}_{2}\,g_{12}
({\bf g}_{12}\cdot {\bf G}_{ij})\ e^{-(\theta_iV_1^2+\theta_jV_2^2)}.
\end{eqnarray}
Here, use has been made of the fact the scalar $\zeta_i^*$ cannot be coupled 
to the traceless tensor $C_{i,k\ell}$ so that the only nonzero contribution 
to the cooling rate comes from the Maxwellian term (first term of the right 
hand side of (\ref{a4.1})) of the distribution function. 
The first integral of Eq.\ (\ref{a5}) is straightforward and can be done
with a change of variables to relative and center of mass variables. The
next two integrals are somewhat more complicated and they can be performed 
by the change of variables 
\begin{equation}
\label{a6}
{\bf x}={\bf V}_1-{\bf V}_2, \quad {\bf y}=\theta_i{\bf V}_1+\theta_j
{\bf V}_2,
\end{equation}
with the Jacobian $(\theta_i+\theta_j)^{-3}$. The integrals can be now 
easily performed and the final result for $\zeta_1$ is 
\begin{eqnarray}
\label{a7}
\zeta_1&=&\frac{2}{3}\sqrt{2}
\left(\frac{\sigma 
_{11}}{\sigma_{12}}\right)^{2}x_1\theta_1^{-1/2}(1-\alpha_{11}^2)\nonumber\\
& & 
+\frac{4}{3}x_2\mu_{21}\left(\frac{\theta_1+\theta_2}{\theta_1\theta_2}
\right)^{1/2}(1+\alpha_{12}) 
\left[2-\mu_{21}(1+\alpha_{12})\frac{\theta_1+\theta_2}{\theta_2}
\right].
\end{eqnarray} 
The result for $\zeta_2$ is obtained from Eq.\ (\ref{a7}) by interchanging 
$1$ and $2$.

\section{Evaluation of the collisional moments}
\label{appB}  

In reduced units, the collisional moments $A_{k\ell}^{ij}$ are given by
\begin{eqnarray}
\label{b1}
A_{k\ell}^{ij}&=&\frac{m_iv_0^2}{T}\pi^{-1/2}\,\int d{\bf V}_1 
V_{1,k}V_{1,\ell}J_{ij}[{\bf V}_1|f_i,f_j]\nonumber\\
&=& \frac{m_iv_0^2}{T}x_j 
\pi^{-1/2}\left(\frac{\sigma_{ij}}{\sigma_{12}}\right)^{2}
\int \,d{\bf V}_{1}\,\int \,d{\bf V}_{2}f_{i}(
{\bf V}_{1})f_{j}({\bf V}_{2}) 
\int d\widehat{\bbox {\sigma}}\,\Theta (\widehat{\bbox {\sigma}}
\cdot {\bf g}_{12})\nonumber\\
& & \times (\widehat{\bbox {\sigma }}\cdot {\bf g}_{12})
\left(V_{1,k}^{''}V_{1,\ell}^{''}-V_{1,k}V_{1,\ell}\right),
\end{eqnarray}
where the identity (\ref{a2}) has been used. Substitution of (\ref{a3}) into 
Eq.\ (\ref{b1}) allows the angular integral to be performed with the result
\begin{eqnarray}
\label{b2}
A_{k\ell}^{ij}&=&-\frac{\sqrt{\pi}}{2} \frac{m_iv_0^2}{T} \mu_{ji}x_j 
\left(\frac{\sigma _{ij}}{\sigma_{12}}\right)^{2}(1+\alpha_{ij})
\int \,d{\bf V}_{1}\,\int \,d{\bf V}_{2}f_{i}(
{\bf V}_{1})f_{j}({\bf V}_{2}) \nonumber\\
& & \times 
\left[g_{12}(G_{ij,k}g_{12,\ell}+G_{ij,\ell}g_{12,k})+\frac{\mu_{ji}}{2}
(3-\alpha_{ij})g_{12}g_{12,k}g_{12,\ell}-\frac{\mu_{ji}}{6}
(1+\alpha_{ij})g_{12}^3\delta_{k\ell}\right],
\end{eqnarray}
where $g_{12,k}=V_{1,k}-V_{2,k}$ and 
$G_{ij,k}=\mu_{ij}V_{1,k}+\mu_{ji}V_{2,k}$. To perform the integral we use
the Sonine approximation of $f_i$ and the change of variables 
(\ref{a6}). When one neglects again nonlinear terms in the tensor ${\sf 
C}_i$, the collisional moment $A_{k\ell}^{ij}$ becomes 
\begin{eqnarray}
\label{b4}
A_{k\ell}^{ij}&=&-\frac{1}{2}\pi^{-5/2} \frac{m_iv_0^2}{T}\mu_{ji}x_j 
\left(\frac{\sigma 
_{ij}}{\sigma_{12}}\right)^{2}(1+\alpha_{ij})\frac{(\theta_i\theta_j)^{3/2}}
{(\theta_i+\theta_j)^3}
\int \,d{\bf x}\,\int \,d{\bf y}e^{-(b_{ij}x^2+d_{ij}y^2)}\nonumber\\
& & \times 
\left[1+\theta_i(\theta_i+\theta_j)^{-2}{\sf C}_i\cdot 
({\bf y}+\theta_j{\bf x})({\bf y}+
\theta_j{\bf x})+\theta_j(\theta_i+\theta_j)^{-2}{\sf C}_{j}\cdot
({\bf y}-\theta_i{\bf x})({\bf y}-
\theta_i{\bf x})\right]\nonumber\\
& & \times
\left[(\theta_i+\theta_j)^{-1}x(x_{k}y_{\ell}+x_{\ell}y_{k})
+\lambda_{ij}xx_kx_{\ell}-\frac{\mu_{ji}}{6}
(1+\alpha_{ij})x^3\delta_{k\ell}\right],
\end{eqnarray}
where 
\begin{equation}
\label{b4.1}
b_{ij}=\theta_i\theta_j(\theta_i+\theta_j)^{-1},
\end{equation}
\begin{equation}
\label{b4.2}
d_{ij}=(\theta_i+\theta_j)^{-1},
\end{equation}
\begin{equation}
\label{b5}
\lambda_{ij}=2\frac{\mu_{ij}\theta_j-\mu_{ji}\theta_i}{\theta_i+\theta_j}+
\frac{\mu_{ji}}{2}(3-\alpha_{ij}).
\end{equation}
The corresponding integrals can be now easily performed and, after some 
algebra, the final result is
\begin{eqnarray}
\label{b5.1}
A_{k\ell}^{ij}&=&\frac{2}{3}\frac{m_iv_0^2}{T}\mu_{ji}x_j 
\left(\frac{\sigma 
_{ij}}{\sigma_{12}}\right)^{2}(1+\alpha_{ij})
\left(\frac{\theta_i+\theta_j}{\theta_i\theta_j}\right)^{3/2} \nonumber\\
& & \times \left\{\left[\case{1}{5}\lambda_{ij}+\case{1}{2}\mu_{ji}
(1+\alpha_{ij})\right]\delta_{k\ell}-2
\frac{\theta_i\theta_j}{(\theta_i+\theta_j)^2}\left[\left(1+\case{3}{5}
\lambda_{ij}\frac{\theta_i+\theta_j}{\theta_i}\right)\gamma_i^{-1}
P_{i,k\ell}\right.\right.\nonumber\\
& & \left.\left.-\left(1-\case{3}{5}
\lambda_{ij}\frac{\theta_i+\theta_j}{\theta_j}\right)\gamma_j^{-1}
P_{j,k\ell}\right]\right\}.
\end{eqnarray}

{}From this general expression one can get the collisional moments 
$A_{k\ell}^{11}$, $A_{k\ell}^{12}$, $A_{k\ell}^{22}$, and 
$A_{k\ell}^{21}$. In particular, 
\begin{equation}
\label{b6}
A_{k\ell}^{11}=F_{11}\delta_{k\ell}+G_{11} P_{1,k\ell},
\end{equation}
\begin{equation}
\label{b7}
A_{k\ell}^{12}=F_{12}\delta_{k\ell}+G_{12} P_{1,k\ell}+H_{12}P_{2,k\ell},
\end{equation}
where
\begin{equation}
\label{b8}
F_{11}=\frac{4}{15}\sqrt{2}\mu_{21}^{-1}x_1\left(\frac{\sigma_{11}}
{\sigma_{12}}\right)^2\theta_1^{-3/2}(1+\alpha_{11})(2+\alpha_{11}),
\end{equation}
\begin{equation}
G_{11}=-\frac{2}{5}\sqrt{2}x_1\left(\frac{\sigma_{11}}{\sigma_{12}}
\right)^2\theta_1^{-1/2}(1+\alpha_{11})(3-\alpha_{11}),
\end{equation}
\begin{equation}
\label{b9}
F_{12}=\frac{4}{3}x_2(1+\alpha_{12})
\left(\frac{\theta_1+\theta_2}{\theta_1\theta_2}\right)^{3/2}\left[
\case{1}{5}\lambda_{12}+\case{1}{2}\mu_{21}(1+\alpha_{12})\right],
\end{equation}
\begin{equation}
\label{b10}
G_{12}=-\frac{8}{3}x_2\mu_{21}(1+\alpha_{12})
\left(\frac{\theta_1}{\theta_2(\theta_1+\theta_2)}\right)^{1/2}\left(
1+\frac{3}{5}\lambda_{12}
\frac{\theta_1+\theta_2}{\theta_1}\right),
\end{equation}
\begin{equation}
\label{b11}
H_{12}=\frac{8}{3}x_2\mu_{12}(1+\alpha_{12})
\left(\frac{\theta_2}{\theta_1(\theta_1+\theta_2)}\right)^{1/2}\left(
1-\frac{3}{5}\lambda_{12}
\frac{\theta_1+\theta_2}{\theta_2}\right).
\end{equation}
The corresponding expressions for $F_{22}$, $G_{22}$, $F_{21}$, $G_{21}$, 
and $H_{21}$ can be easily inferred from Eqs.\ (\ref{b8})--(\ref{b11}) by 
just making the changes $1\leftrightarrow2$. From Eqs.\ 
(\ref{b6})--(\ref{b11}) and (\ref{a7}), it is easy to prove the identity 
\begin{equation}
\label{b13}
-\gamma_1\zeta_1=F_{11}+F_{12}+\left(G_{11}+G_{12}\right)\gamma_1+
H_{12}\gamma_2,
\end{equation}
which is in fact required by the partial energy conservation 
equation (\ref{2.19}) to support the solution found for the simple shear 
flow problem. In the case of mechanically 
equivalent particles, the expression of the collisional moments 
$A_{k\ell}^{ij}$ are
\begin{equation}
\label{b12}
A_{k\ell}^{11}+A_{k\ell}^{12}=\frac{4}{15}(1+\alpha)\left[(2+\alpha)
\delta_{k\ell}-\frac{3}{2}(3-\alpha)P_{k\ell}\right].
\end{equation} 
When $\alpha=1$, Eq.\ (\ref{b12}) reduces to the results derived from the 
Boltzmann equation in the first Sonine approximation.

\section{Low dissipation limit}
\label{appC}  

In this Appendix we derive the expressions for the main quantities of the 
simple shear flow problem in the low-dissipation limit. The expansions in 
powers of $\epsilon\equiv(1-\alpha^2)^{1/2}$ are given by Eqs.\ 
(\ref{2.34})--(\ref{2.38}). For symmetry reasons, the expansion of 
$P_{i,xy}^*$ has only odd powers, while those of the normal stresses (and 
of the temperature ratio) have only even powers. However, from a practical 
point of view, it is simpler to use $a^*\equiv a/\nu$ as a perturbation 
parameter instead of $\epsilon$. Thus, the expansions 
(\ref{2.34})--(\ref{2.38}) can be written in terms of $a^*$ as
\begin{equation}
\label{c1}
\alpha=1+\alpha_0 a^{*2}+\cdots,
\end{equation}
\begin{equation}
\label{c2}
\gamma=1+\gamma_0 a^{*2}+\cdots,
\end{equation}
\begin{equation}
\label{c3}
P_{i,yy}^*=1+P_{i,yy}^{(2)}a^{*2}+\cdots,
\end{equation}
\begin{equation}
\label{c4}
P_{i,xx}^*=1+P_{i,xx}^{(2)} a^{*2}+\cdots,
\end{equation}
\begin{equation}
\label{c5}
P_{i,xy}^*=P_{i,xy}^{(1)}a^{*}+\cdots.
\end{equation}
Of course, both expansions are directly related, so that 
\begin{equation}
\label{c6}
a_0=\frac{1}{\sqrt{-2\alpha_0}}.
\end{equation}

The coefficients $P_{i,k\ell}^{(r)}$ can be obtained from Eq.\ (\ref{2.27}) 
by retaining terms up to second order in $a^*$. To do that, we need the 
corresponding expansions of the quantities $F_{ij}$, $G_{ij}$, and $H_{ij}$. 
They are given by
\begin{eqnarray}
\label{c7}
F_{11}&=&\frac{8}{5}\sqrt{2\mu_{21}}x_1\left(\frac{\sigma_{11}}
{\sigma_{12}}\right)^2\left[1+(\case{5}{6}\alpha_0-\case{3}{2}
x_2\gamma_0)a^{*2}+\cdots \right]
\nonumber\\
&\equiv&F_{11}^{(0)}+F_{11}^{(2)}a^{*2}+\cdots,
\end{eqnarray}
\begin{eqnarray}
\label{c8}
G_{11}&=&-\frac{8}{5}\sqrt{2\mu_{21}}x_1\left(\frac{\sigma_{11}}
{\sigma_{12}}\right)^2\left(1+\case{1}{2}\gamma_0x_2a^{*2}+\cdots 
\right)
\nonumber\\
&\equiv&G_{11}^{(0)}+G_{11}^{(2)}a^{*2}+\cdots,
\end{eqnarray} 
\begin{eqnarray}
\label{c9}
F_{12}&=&\frac{16}{5}\mu_{21}x_2
\left(1+\frac{5\alpha_0+\gamma_0(9x_2-7\mu_{12})}{6} a^{*2}+\cdots\right)
\nonumber\\
&\equiv&F_{12}^{(0)}+F_{12}^{(2)}a^{*2}+\cdots,
\end{eqnarray}
\begin{eqnarray}
\label{c10}
G_{12}&=&-\frac{16}{15}x_2\mu_{21}(3+2\mu_{12})
\left\{1+\frac{5\alpha_0\mu_{12}
-\gamma_0\left[4\mu_{12}^2+\mu_{12}(2x_1-1)+3(x_1-1)\right]}
{2(3+2\mu_{12})}a^{*2}+\cdots\right\}
\nonumber\\
&\equiv&G_{12}^{(0)}+G_{12}^{(2)}a^{*2}+\cdots,
\end{eqnarray}
\begin{eqnarray}
\label{c11}
H_{12}&=&\frac{16}{15}x_2\mu_{12}
\left[1+\frac{3}{2}\mu_{21}(\alpha_0-2\gamma_0\mu_{12})a^{*2}+\cdots\right]
\nonumber\\
&\equiv&H_{12}^{(0)}+H_{12}^{(2)}a^{*2}+\cdots.
\end{eqnarray}
The expressions of the coefficients $F_{ij}^{(r)}$, $G_{ij}^{(r)}$, and 
$H_{ij}^{(r)}$ can be easily identified from the above equations. On the 
other hand, the coefficients $F_{22}$, $G_{22}$, $F_{21}$, $G_{21}$, and 
$H_{21}$ are obtained by just making the changes $1\leftrightarrow 2$, 
$\alpha_0\to \alpha_0$ and $\gamma_0\to -\gamma_0$. Substitution of Eqs.\ 
(\ref{c7})--(\ref{c11}) (and their counterparts for species 2) into Eq.\ 
(\ref{2.27}) and considering only terms through second order in $a^*$ allows 
one to get the coefficients $P_{i,k\ell}^{(r)}$ in terms of $\alpha_0$ and 
$\gamma_0$. The final expressions of such coefficients will be omitted here 
since they are very large and not very illuminating.

Once the pressure tensors are known, we are in conditions to get the 
coefficients $\alpha_0$ and $\gamma_0$. First, the cooling rates behave as
\begin{eqnarray}
\label{c12}
\zeta_1^*&=&-\frac{4}{3}\left[\left(\sqrt{2\mu_{21}}\left(\frac{\sigma_{11}}
{\sigma_{12}}\right)^2x_1+2\mu_{21}x_2\right)\alpha_0-4\mu_{21}\mu_{12}x_2
\gamma_0\right]a^{*2}+\cdots\nonumber\\
&\equiv& 
\left(A_1\alpha_0+B_1\gamma_0\right)a^{*2}+\cdots\;,
\end{eqnarray}
\begin{eqnarray}
\label{c13}
\zeta_2^*&=&-\frac{4}{3}\left[\left(\sqrt{2\mu_{12}}\left(\frac{\sigma_{22}}
{\sigma_{12}}\right)^2x_2+2\mu_{12}x_1\right)\alpha_0+4\mu_{12}\mu_{21}x_1
\gamma_0\right]a^{*2}+\cdots\nonumber\\
&\equiv& 
\left(A_2\alpha_0+B_2\gamma_0\right)a^{*2}+\cdots\;,
\end{eqnarray}
where again the coefficients $A_i$ and $B_i$ are easily identified. The 
quantities $\alpha_0$ and $\gamma_0$ can be now easily obtained from the 
partial balance of energy (\ref{2.19}), i.e.,
\begin{equation}
\label{c14}
-\frac{2}{3}P_{i,xy}^{(1)}=A_i\alpha_0+B_i\gamma_0 \quad (i=1,2).
\end{equation}
The solution of this system of algebraic equations leads to 
\begin{equation}
\label{c15}
\alpha_0=\frac{2}{3}\frac{B_1 P_{2,xy}^{(1)}-
B_2 P_{1,xy}^{(1)}}{A_1B_2-A_2B_1},
\end{equation}
\begin{equation}
\label{c16}
\gamma_0=\frac{2}{3}\frac{A_2 P_{1,xy}^{(1)}-
A_1P_{2,xy}^{(1)}}{A_1B_2-A_2B_1},
\end{equation}
where 
\begin{equation}
\label{c17}
P_{1,xy}^{(1)}=\frac{H_{12}^{(0)}(F_{22}^{(0)}+F_{21}^{(0)})(G_{11}^{(0)}+
G_{22}^{(0)}+G_{12}^{(0)}+G_{21}^{(0)})-(F_{11}^{(0)}+F_{12}^{(0)})\left[
(G_{22}^{(0)}+G_{21}^{(0)})^2+H_{12}^{(0)}H_{21}^{(0)}\right]}{
\left[(G_{11}^{(0)}+G_{12}^{(0)})(G_{22}^{(0)}+G_{21}^{(0)})-H_{12}^{(0)}
H_{21}^{(0)}\right]^2},
\end{equation}
and $P_{2,xy}^{(1)}$ can be obtained from $P_{1,xy}^{(1)}$ by setting the 
changes $1\leftrightarrow 2$. 

In summary, Eqs.\ (\ref{c15}) and (\ref{c16}) give $\alpha_0$ and 
$\gamma_0$, respectively, while $a_0$ is given by (\ref{c6}) 
and the coefficients $P_{i,k\ell}^{(r)}$ are obtained from Eq.\ (\ref{2.27}) 
when only terms through second order in the dissipation parameter 
$\epsilon$ are retained.

\begin{figure}
\caption{Time evolution of the reduced granular temperature 
$T^*(t)=\nu^2(t)/a^2$ as obtained from Monte Carlo simulation of the 
Boltzmann equation for $\alpha=0.75$, $\protect\sigma_{11}= 
\protect\sigma_{22}$, $n_1/n_2=1/3$, $m_1/m_2=4$, and starting from three 
different initial conditions. Time is measured in units of 
$\lambda_{11}/V_{01}(0)$.}
\label{fig1}
\end{figure}

\begin{figure}
\caption{Plot of the reduced granular temperature $T^*=\nu^2/a^2$ versus 
the parameter $(1-\alpha^2)^{-1}$ as obtained from simulation (symbols) and 
the Sonine approximation (lines), for $w=\sigma_{11}/\sigma_{22}=1$, 
$\delta=n_1/n_2=1$ and three different values of the mass ratio 
$\mu=m_1/m_2$: $\mu=10$ (solid line and circles); $\mu=2$ (dashed line and 
squares), and $\mu=1$ (dotted line and triangles).}
\label{fig2}
\end{figure}

\begin{figure} 
\caption{Plot of the temperature ratio $\gamma=T_1/T_2$ as a function 
of the restitution coefficient $\alpha$ as obtained from simulation 
(symbols) and the Sonine approximation (lines). We have considered  
$w=\sigma_{11}/\sigma_{22}=1$, $\delta=n_1/n_2=1$ and three different values 
of the mass ratio $\mu=m_1/m_2$: $\mu=10$ (solid line and circles); $\mu=2$ 
(dashed line and squares), and $\mu=1$ (dotted line and triangles).}
\label{fig3}
\end{figure}

\begin{figure} 
\caption{Plot of the reduced elements of the pressure tensor (a) 
$-P_{xy}^*=-P_{xy}/p$, (b) $P_{xx}^*=P_{xx}/p$, (c) $P_{yy}^*=P_{yy}/p$ and 
(d) $P_{zz}^*=P_{zz}/p$ versus the restitution coefficient $\alpha$ as 
obtained from simulation (symbols) and the Sonine approximation (lines). We 
have considered  $w=\sigma_{11}/\sigma_{22}=1$, $\delta=n_1/n_2=1$ and three 
different values of the mass ratio $\mu=m_1/m_2$: $\mu=10$ (solid line and 
circles); $\mu=2$ (dashed line and squares), and $\mu=1$ (dotted line and 
triangles).}
\label{fig4}
\end{figure}

\begin{figure}
\caption{Plot of the temperature ratio $\gamma=T_1/T_2$ as a function of the 
restitution coefficient $\alpha$ as obtained from simulation (symbols) and 
the Sonine approximation (lines). We have considered  
$w=\sigma_{11}/\sigma_{22}=1$, $\mu=m_1/m_2=4$ and two different values of 
the concentration ratio $\delta=n_1/n_2$: $\delta=1/3$ (solid line and 
circles), and $\delta=3$ (dashed line and squares).}
\label{fig5}
\end{figure}

\begin{figure}
\caption{Plot of the reduced elements of the pressure tensor (a) 
$-P_{xy}^*=-P_{xy}/p$, (b) $P_{xx}^*=P_{xx}/p$, (c) $P_{yy}^*=P_{yy}/p$ and 
(d) $P_{zz}^*=P_{zz}/p$ versus the restitution coefficient $\alpha$ as 
obtained from simulation (symbols) and the Sonine approximation (lines). We 
have considered  $w=\sigma_{11}/\sigma_{22}=1$, $\mu=m_1/m_2=4$ and two 
different values of the concentration ratio $\delta=n_1/n_2$: $\delta=1/3$ 
(solid line and circles), and $\delta=3$ (dashed line and squares).}
\label{fig6}
\end{figure}

\begin{figure}
\caption{Plot of the reduced elements of the pressure tensor (a) 
$-P_{xy}^*=-P_{xy}/p$, (b) $P_{xx}^*=P_{xx}/p$, (c) $P_{yy}^*=P_{yy}/p$ and 
(d) $P_{zz}^*=P_{zz}/p$ versus the restitution coefficient $\alpha$ as 
obtained from simulation (symbols) and the Sonine approximation (lines). We 
have considered  $\mu=m_1/m_2=1$, $\delta=n_1/n_2=1$, and two 
different values of the size ratio $w=\sigma_{11}/\sigma_{22}$: $w=2$ 
(solid line and circles), and $w=1$ (dashed line and triangles).}
\label{fig7}
\end{figure}


\begin{references}
\bibitem{note}As a matter of fact, the temperature differences found in Ref.
\onlinecite{JM87} are due to effects of finite density (effects of
excluded area) and so vanish in the low-density limit.
\end{references}
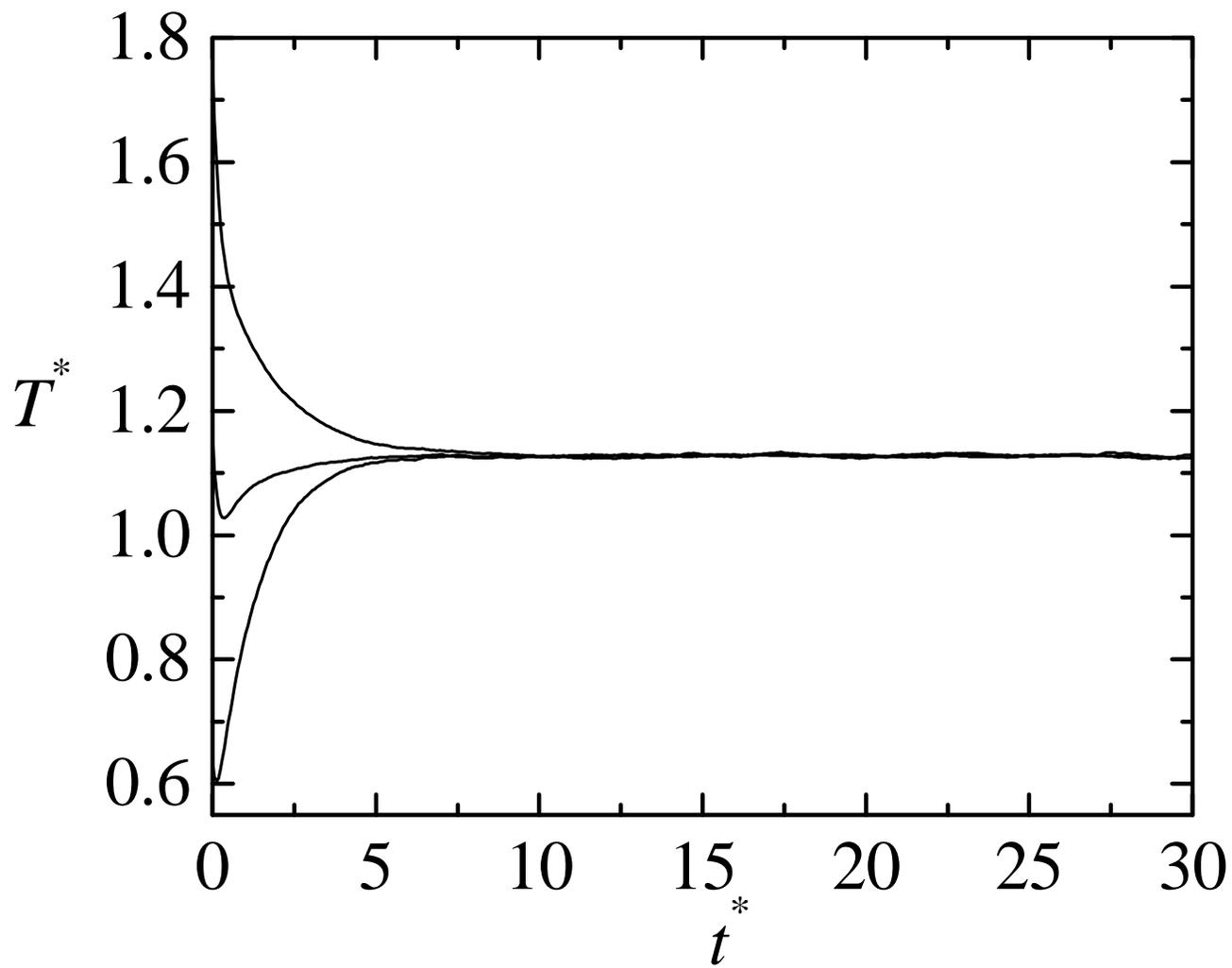
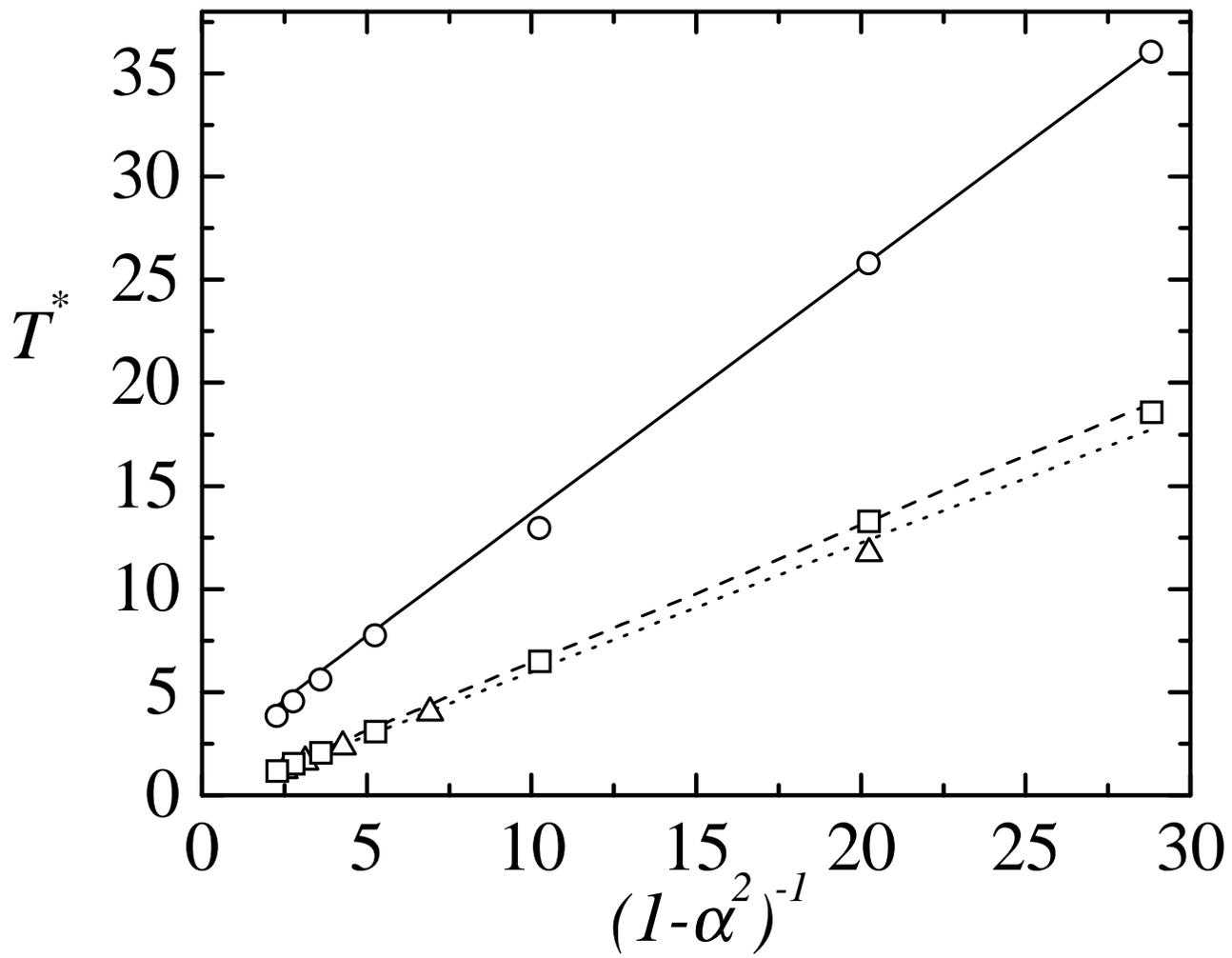
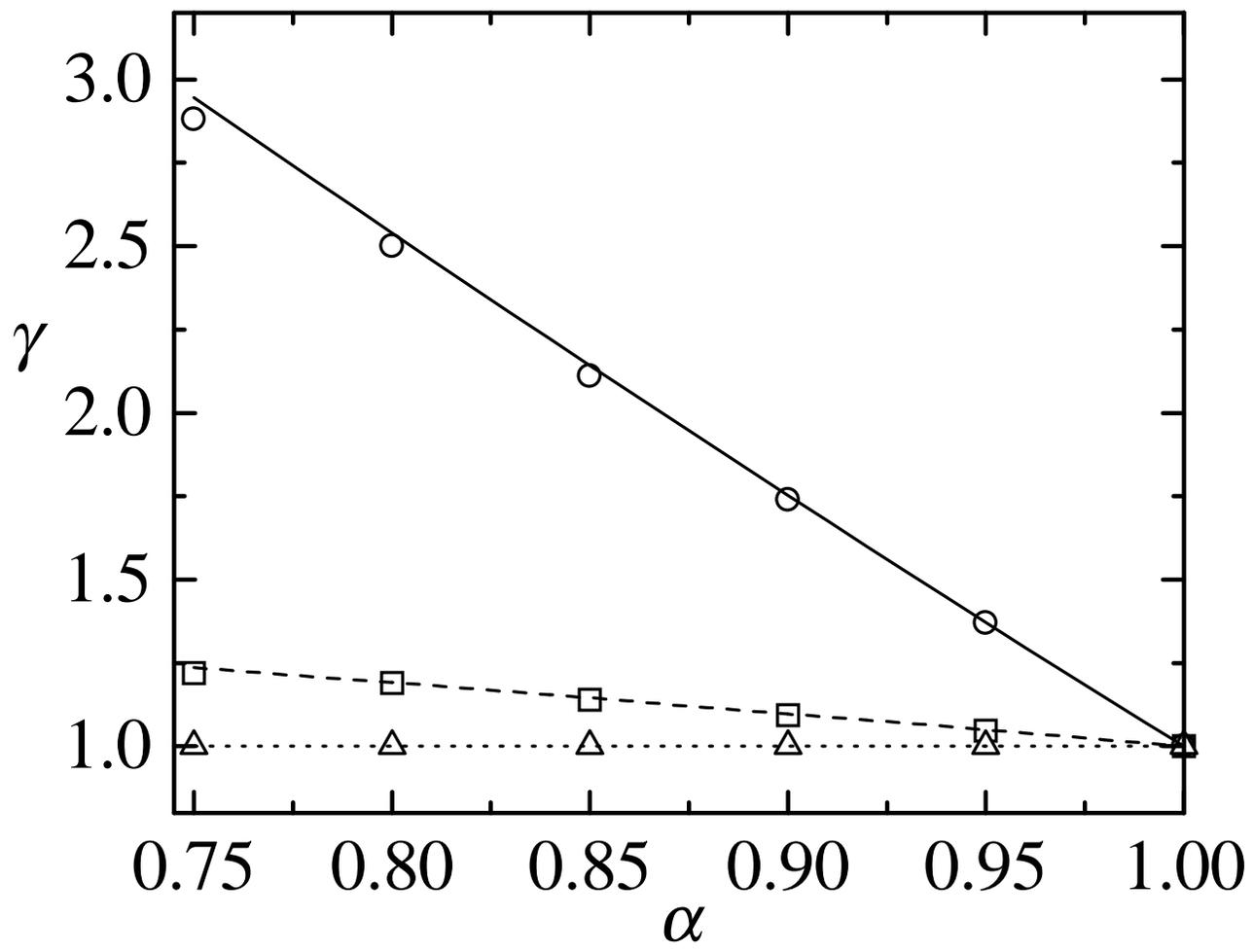
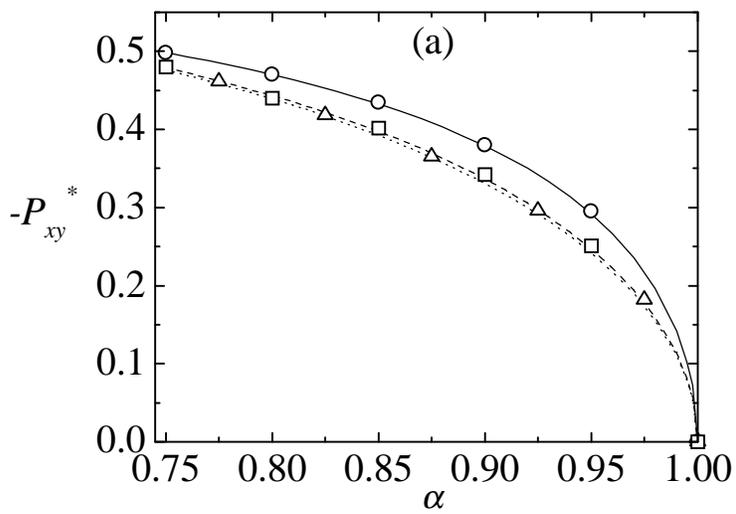
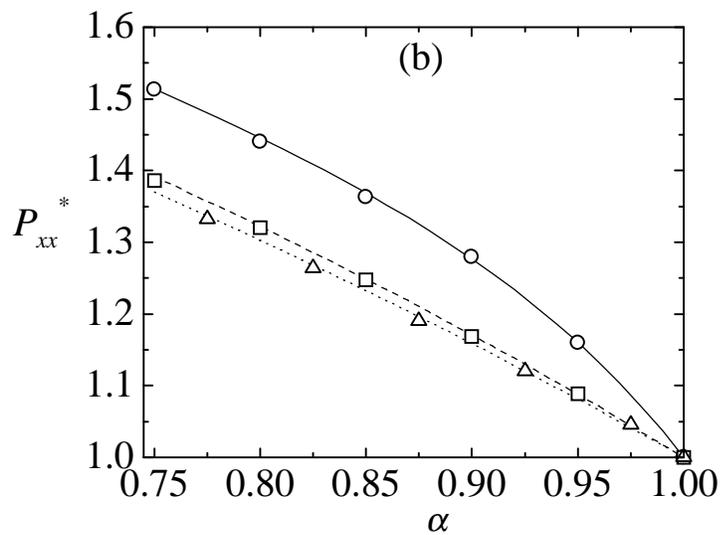
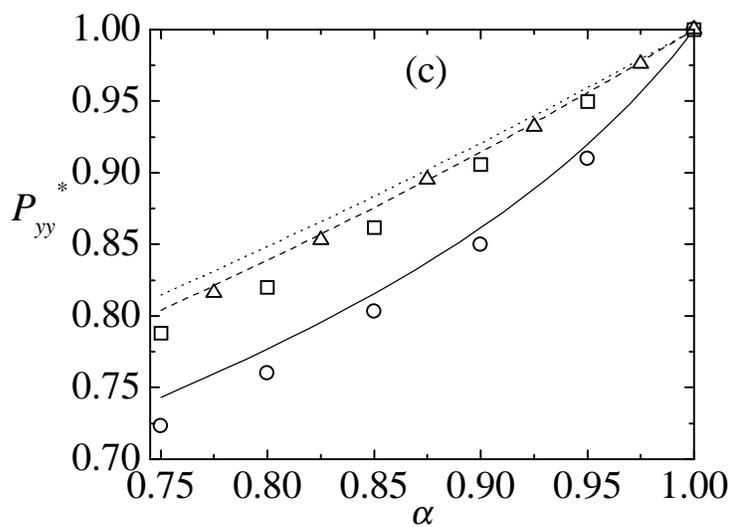
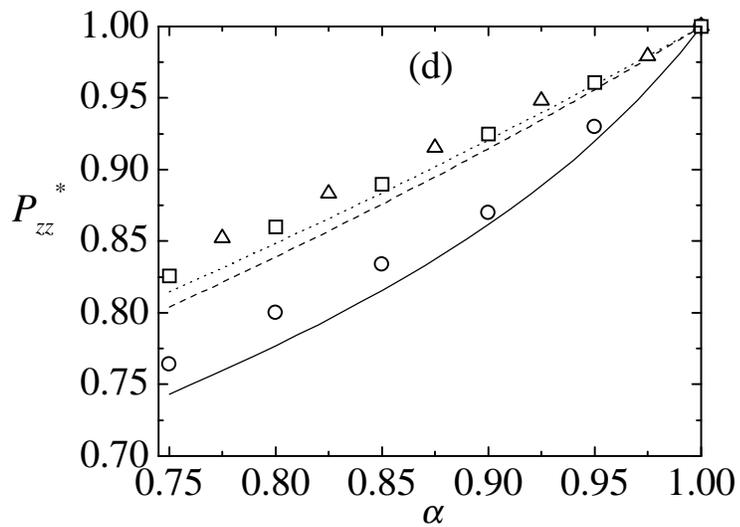
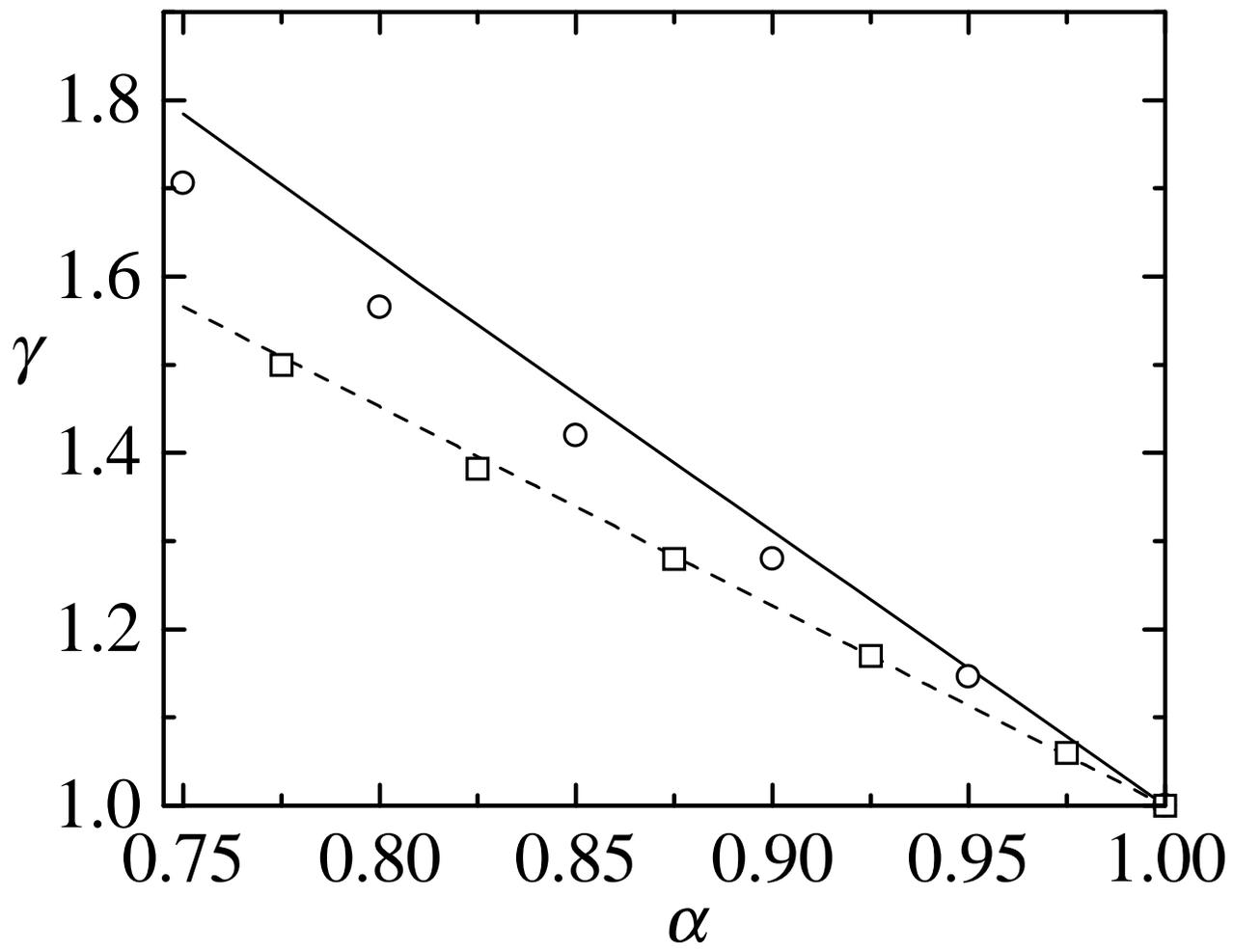
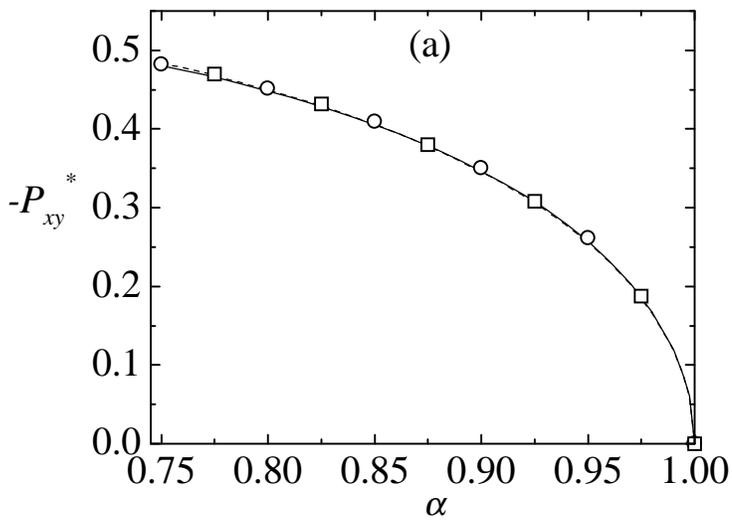
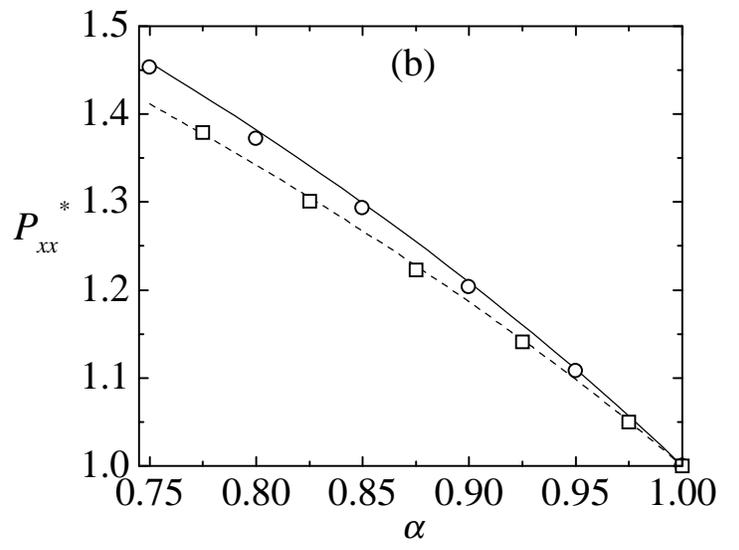
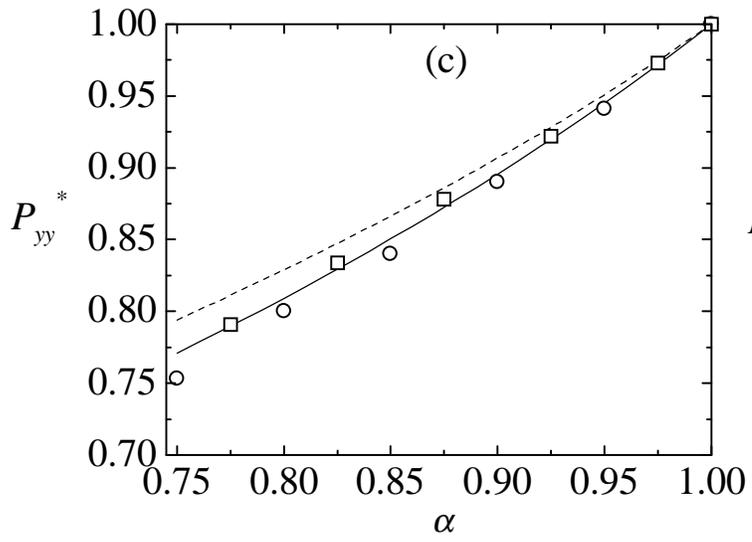
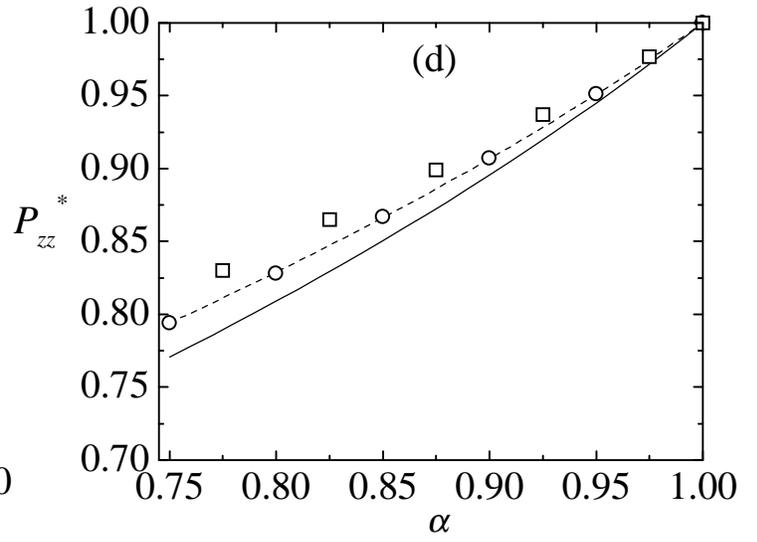
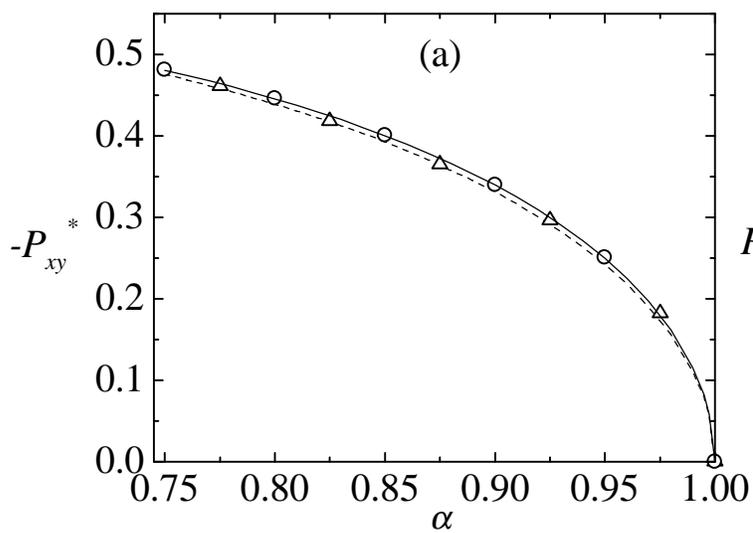
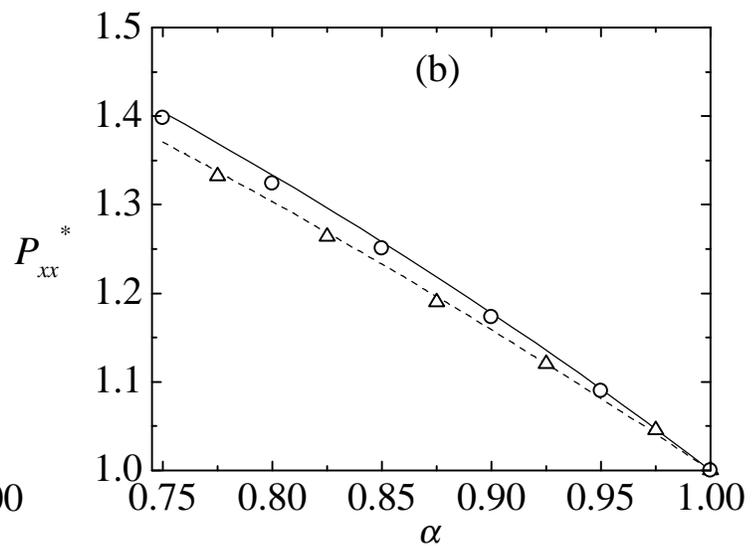
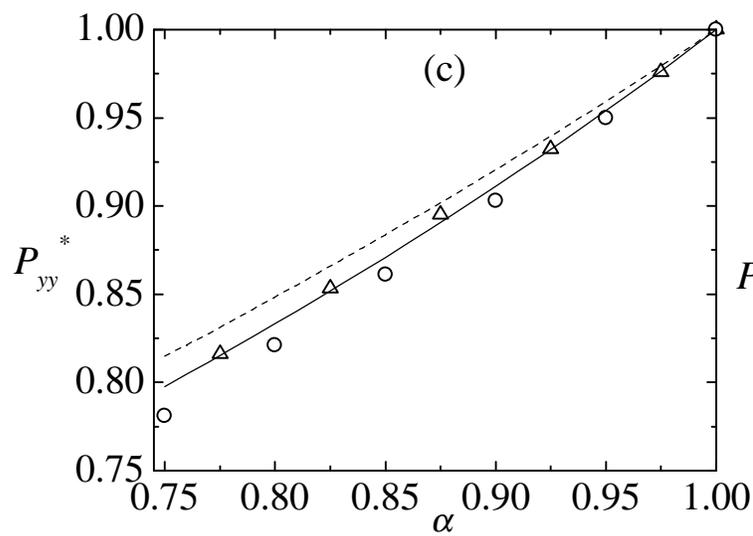
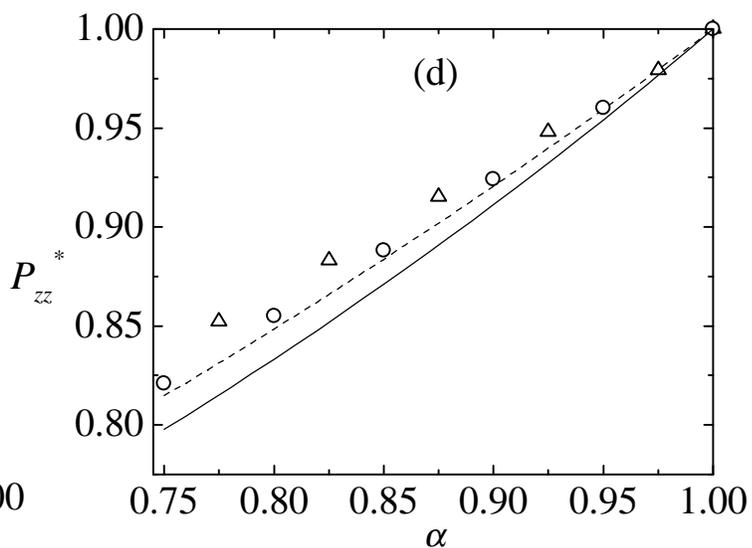
\end{document}